\documentclass[12pt]{iopart}
\usepackage{graphicx}
\usepackage{bm}
\usepackage{amstext}

\newcommand{\thc}[1]{\multicolumn{1}{c}{\textbf{#1}}}
\newcommand{\thl}[1]{\multicolumn{1}{l}{\textbf{#1}}}
\renewcommand{\deg}[0]{$^{\circ}$}

\begin{document}

\title[H$_2$O Interaction Potential Based on a Single Center
Multipole Expansion]{A Transferable H$_2$O Interaction Potential Based on a Single Center
Multipole Expansion: SCME}

\author{K T Wikfeldt$^1$, E R Batista$^2$, F D Vila$^3$ and H J\'onsson$^4$}
\address{$^1$ Science Institute of the University of Iceland, VR-III, 107, Reykjav\'{i}k, Iceland and Nordita, 
KTH Royal Institute of Technology and Stockholm University Roslagstullsbacken 23, SE 106 91 Stockholm, Sweden
$^2$ Department of Chemistry, University of Washington, Seattle, WA 98195. 
\footnote{Present address: Theoretical Division, T-12 MS B268, Los Alamos National Laboratory, Los Alamos, NM 87545}
$^3$ Department of Chemistry, University of Washington, Seattle, WA 98195.
\footnote{Present address: Department of Physics, University of Washington, Seattle, WA 98195}
$^4$ Department of Chemistry, University of Washington, Seattle, WA 98195 and
Faculty of Physical Sciences, VR-III, University of Iceland, 107, Reykjav\'{i}k, Iceland
}

\ \ \ 

\vskip 1 true cm


\begin{abstract}
A transferable potential energy function for describing the interaction between
water molecules is presented. The electrostatic interaction is described rigorously 
using a multipole expansion. Only one expansion center is used per molecule to 
avoid the introduction of monopoles.  This single center approach turns out to 
converge and give close agreement with ab initio calculations when carried out up 
to and including the hexadecapole.  Both dipole and quadrupole polarizability is 
included. All parameters in the electrostatic interaction as well as the dispersion 
interaction are taken from \textit{ab initio} calculations or 
experimental measurements of a single water molecule. The repulsive part of the 
interaction is parametrized to fit \textit{ab initio} calculations of small water 
clusters and experimental measurements of ice I$_h$. 
The parametrized potential function was then used to simulate liquid water and the results
agree well with experiment, even better than simulations using some of the point 
charge potentials fitted to liquid water. 
The evaluation of the new interaction potential for 
condensed phases is fast because point charges are not present and the 
interaction can, to a good approximation, be truncated at a finite range.
\end{abstract}


\maketitle

\section{Introduction}
\label{sec:intro}

Water in its various forms plays a fundamental role in many
biological, chemical and physical processes\cite{ball2001}.
Hydration water around biomolecules participates actively in
biological function such as protein folding~\cite{cheung2002}, and the complex
interactions between biomolecules inside cells is mediated by the
water solvent through the hydrophobic effect\cite{li2011,snyder2011,grossman2011}.
Supercooled water in the bulk and in confined geometries 
is also of large current interest due to the intriguing yet controversial possibility 
of a liquid-liquid critical point in the deeply supercooled 
region\cite{speedy1976,poole1992,debenedetti2003}.
On a larger scale, global climate change is affected by feedback
loops involving water vapor \textemdash\ the most common greenhouse gas \textemdash\ and
liquid water \cite{pierrehumbert99,held00}.
Moreover, our environment depends critically on the properties of ice
\cite{petrenko99a,smith02}, both through the rheology of ice 
sheets\cite{paterson94}
and the meteorology of clouds\cite{pruppacher97}.
Ice is also found in interstellar space, where, in an
amorphous phase, it coats dust grains in molecular 
clouds\cite{ehrenfreund02,ehrenfreund00}. These coatings can serve as a 
substrate for the formation of chemicals of biological
interest\cite{caro02}. In spite
of the large amounts of information available, the molecular mechanisms
behind all of these processes 
are just beginning to be understood.

The water molecules involved in the most common processes in nature are in an environment that is
characteristic of neither liquid water, ice nor water vapor, e.g.
amorphous ice\cite{burton35,mayer82,mccammon1985,jenniskens94,hallbrucker89,debenedetti2003},
premelted\cite{dash95,furukawa97,kroes92}
and solid\cite{morgenstern97,materer97,braun98} surfaces and adsorbed overlayers.
The correct description of such systems
is in many cases beyond the computational capabilities of available
\textit{ab initio} methods. Nowadays most condensed phase systems are studied
by means of density functional theory (DFT)\cite{kryachko90,march96} or
model potentials\cite{frenkel02}. In the case
of water, however, DFT methods are handicapped by both theoretical and practical
reasons\cite{stone96a}: first, the results obtained for systems containing hydrogen bonds
are rather mixed \cite{xantheas95,delbene95,sprik96,grossman04,ireta2004,santra2007}. 
Secondly, the most commonly used functionals do not correctly account for 
the long-range $R^{-6}$ terms corresponding
to the dispersion energy, and are therefore unable to correctly model weak
intermolecular interactions\cite{perez95,vanmourik02}. A new class of so called
vdW functionals that include a 
description of non-local interactions have been 
introduced\cite{dion04,vydrov09,lee10,klimes10}, 
but their accuracy is still subject to debate
and for many applications the computational demands are
too high.

Interaction potential functions, on the other
hand, usually have  low computational
requirements and have been successful in modeling various
aspects of water\cite{wallqvist99,vega11}. 
The functions most commonly used are simple
two-body effective potentials such as SPC/E\cite{berendsen87},
TIP3P\cite{jorgensen83} and TIP4P\cite{jorgensen83} (and more recently 
improved reparametrizations such as TIP4P/Ew\cite{horn04} and 
TIP4P/2005\cite{abascal05}) which were developed
to reproduce the 
structural and thermodynamic properties of bulk
phases at ambient temperature and pressure. 
A common feature of these potentials is enhanced
multipole moments of the molecules representing the effects 
of the mean-field, many-body polarization seen in the liquid and 
the solid. Although this approach gives reasonable results for
several properties of 
the bulk phase, it has been shown that the explicit introduction of many-body 
polarization effects is required to accurately describe other environments, 
for example water clusters\cite{pedulla98,mhin93,wales98,dang97}.
Pedulla and Jordan\cite{pedulla98} have shown that non-additive interactions 
play an important role in the description of
phase changes in small clusters, an observation that is likely to extend to processes
such as premelting, island formation on surfaces and diffusion. Polarizable model
potentials such as NCC\cite{niesar90} and DC\cite{dang97} have been shown to give good results
for both small
clusters and the liquid, and modifications of the DC potential provide an acceptable
description of ice\cite{dong01}. More recently Millot \textit{et al.}\cite{millot92,millot98}
and Burnham and Xantheas\cite{burnham02a,burnham02b,fanourgakis08} have 
presented transferable potentials that reproduce well
\textit{ab initio} results for clusters.

An important concern when modeling condensed phases is long range interactions, 
i.e. the interaction between atoms and molecules
separated by large distance.
The contribution of such long range interactions, beyond a cutoff radius of $R_c$,
to the energy of the system can be obtained by integration as
\begin{equation}
\label{eq:utail}
  U^{\text{tail}}(R_c) ~\propto~ \int_{R_c}^{\infty} u(R)~ 4 \pi R^2~ dR ,
\end{equation}
where $u(R)$ is the interaction potential function. If the potential decays faster than
$R^{-3}$ a value for $R_c$ can 
be determined in such a way that the long range contribution
becomes insignificant and only interactions for distances smaller than $R_c$ need to be included. 
The vast majority of empirical water potentials functions, however, make use point or diffuse
charges on atomic or pseudo-atomic sites, resulting in an interaction between sites that decays
as $R^{-1}$. 
The contribution of this long range tail then diverges and its
effects must be accounted for explicitly. Several methods have been developed for this purpose,
varying in their rigor and computational effort, and their relative merits have been the
subject of much debate.
The most widely used approaches, such as Ewald sums\cite{frenkel02a,ewald21} and
reaction field methods, add a significant computational effort.
Moreover, the use of periodic boundary conditions in the case of the Ewald method
might introduce artificial periodicity effects such as dynamic correlations between images.
The simplest procedure, i.e. truncation of the long-range interactions
due to the point or diffuse charges, is known to 
result in spurious behavior at the cutoff distance\cite{allen87a}.

The widespread use of point charges in model potentials has been a matter of convenience
rather than necessity since the leading term in the electrostatic multipole expansion for a water molecule is
the dipole and the long range interaction consequently decays as $R^{-3}$.
Therefore, the integral in Equation~\ref{eq:utail} 
can converge in certain 
cases for a model potential that avoids point or diffuse charges. 
Two systems of special interest for which such a truncation scheme should 
be feasible are proton
disordered crystals, and surfaces. In the former the long-range interactions tend
to cancel out due to the random orientation of the molecular dipoles, while for surfaces the
volume integral in Equation~\ref{eq:utail} becomes two-dimensional and converges
unconditionally.
The use of charge free potentials is not new. Dipolar fluids are commonly simulated using
Stockmayer-type potentials composed of a Lennard-Jones interaction supplemented with
an embedded point dipole moment. An example of this approach is the "soft sticky dipole"
model of Liu and Ichiye\cite{ichiye96}. These potentials suffer from the
drawback that they are parametrized
to reproduce average properties of bulk water and, for the most part, are not polarizable
and, therefore, not transferable.
A different
approach is the so-called polarizable electropole of Barnes
\textit{et al.}\cite{barnes79} involving a simple approximation to the multipole expansion
based on polarizable dipoles and quadrupoles.
This potential is, however, not of high accuracy and has not been used much.

Previous studies of
a charge free, single-center multipole expansion for the water monomer\cite{batista98,batista00}
have shown that an accurate description of the electric fields in ice and around water clusters
is obtained if the expansion is carried out up to and including the hexadecapole. Due
to the proton-disordered nature of ice Ih, the local electric
field at a water molecule due to its surroundings was shown to be converged for a cutoff radius of only
8 \AA \cite{batista98}. This approach has several
advantages over the distributed multipole expansion\cite{millot92,millot98}, where two or more centers of a multipole expansion are placed on each molecule.
For example, the use of a single center requires significantly less computational effort in the
iterative solution of the polarization equations. 
Secondly, since no point charges are present and the long range interaction therefore decays quickly,  it is possible to introduce a finite range cutoff, $R_c$,
and avoid the computationally demanding Ewald summation.

In the present article, we extend these studies and present a complete model
potential function where the electrostatic
and induction parameters are obtained for a single water molecule, thus allowing the
condensed phase properties to emerge from the molecular properties through polarizability and self-consistent calculations of the local field.
This construction of the potential function ensures transferability to
different kinds of environments, while the truncation of long range interactions makes it easier to carry out long simulations on complex systems. 
The goal is to create a potential energy function that reproduces accurate
\textit{ab initio} calculations of the Born-Oppenheimer potential surface.
Quantum mechanical effects such as zero-point energy are not
built into the potential, unlike for example the SPC/E and TIP4P potentials where the fitting to
experimental data indirectly brings in some average quantum mechanical effects, appropriate only for bulk water at ambient conditions. 
In the following section we describe the
different
components 
of the potential
in detail, as well as
the various procedures used to obtain the parameters involved. Section
\ref{sec:resul} presents and discusses the results for the (H$_2$O)$_{n}$
clusters with $n=2$ to $6$ (with special emphasis on the dimer), liquid water and ice Ih, the most
common crystal structure of ice. Finally, Section \ref{sec:concl} presents conclusions
and future perspectives.


\section{Definition of the Potential Function}
\label{sec:metho}

The vast majority of interaction potentials are based in one way or another
on the long- and short-range perturbation theories of intermolecular interactions
\cite{stone96b}.
The former applies when the separation between molecules is sufficiently large for 
the overlap between wave functions 
to be insignificant. In such a case the exact expression
for the interaction energy reduces to a sum of electrostatic, induction and
dispersion terms. At shorter distances, however, the exchange repulsion
and in some cases the charge-transfer arising from the overlap cannot be ignored. Since the
evaluation of the interaction at intermediate and short range is difficult,
the electrostatic, induction and
dispersion terms arising from the long-range perturbation theory are often simply scaled by
means of damping functions 
at short range
and complemented by a short-range repulsion\cite{ahlrichs77}.
With the exception of the ASP family of model interaction
potentials\cite{millot92,millot98}, the charge-transfer component is not explicitly
included and is usually folded into the other components through the parametrization, a
simplification which is justified due to the small magnitude of this effect\cite{stone96c}.

\begin{figure*}
\includegraphics[scale=0.15]{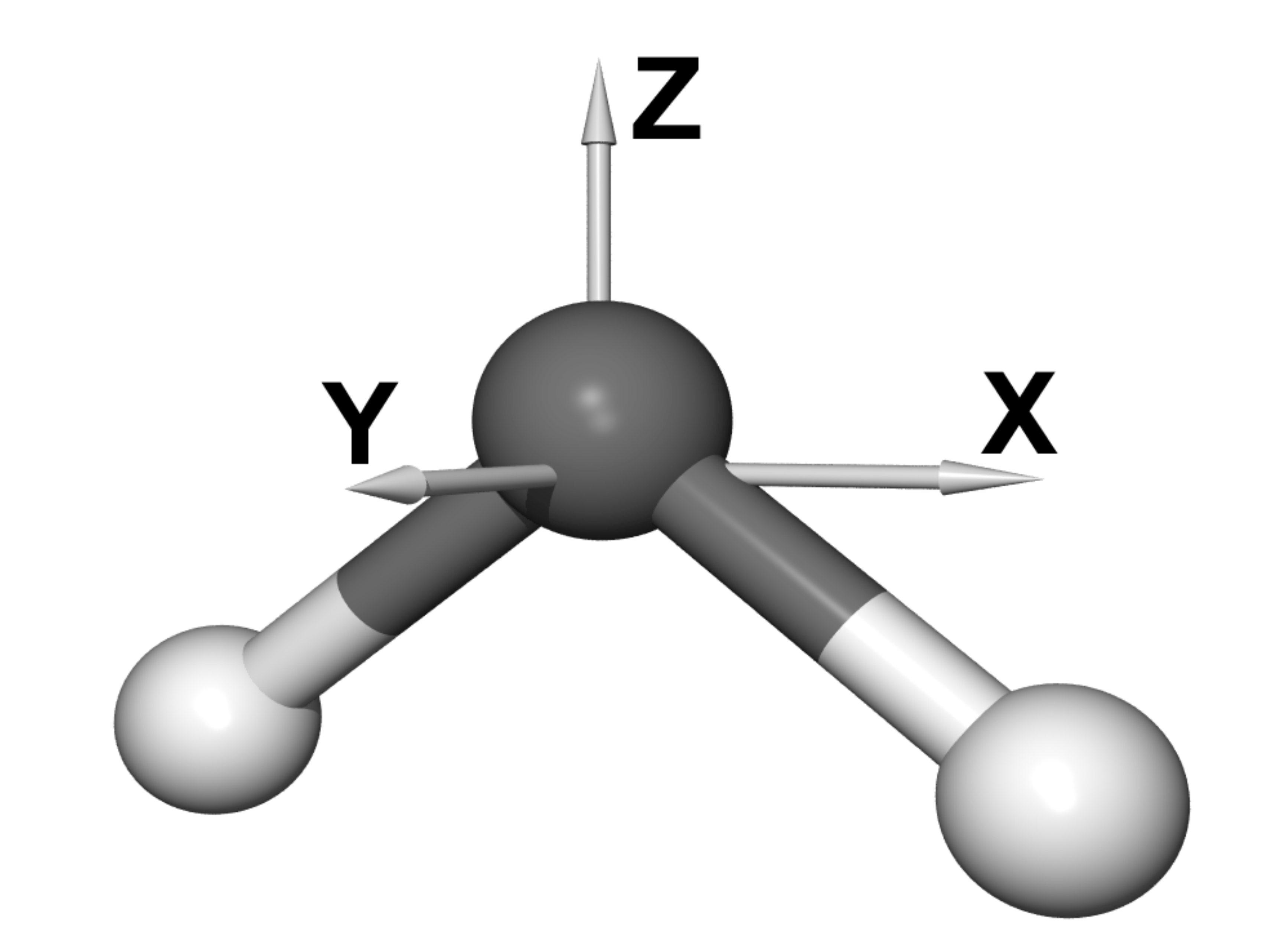}
\caption{\label{fig:w1axes}
         Molecular Cartesian coordinate system with origin in the center of mass
         used in the definition of the multipole moments and polarizabilities.
        }
\end{figure*}

Following this 
approach, we have defined the total interaction energy
between water molecules as the sum of electrostatic, induction, dispersion and
short-range repulsion terms:
\begin{equation}
  E_{\text{tot}} ~=~ E_{\text{es+ind}} + E_{\text{disp}} + E_{\text{rep}}~.
\end{equation}
Each water molecule is treated as a rigid body with fixed bond length and bond angle.
We have chosen the experimentally determined molecular conformation
($r_{\text{OH}}=0.9572 \text{\AA}$, $\widehat{\text{HOH}}=104.52$\deg)
to define the center of mass, but the interaction potential presented here is independent
of that choice. A Cartesian coordinate system with origin on the center of mass is defined
as shown in Figure \ref{fig:w1axes}. The center of mass was used as a reference point
in the calculation of the of electrostatic and 
induction
components. The other components, i.e. the dispersion and repulsion, are naturally centered
on the oxygen atom. Two auxiliary centers are used simply to orient the multipole moments
associated with each monomer and are located on the hydrogen atoms.

\subsection{Electrostatic and Induction Energies}
\label{subsec:esindcom}
The electric interaction between the molecules is described in
terms of a single-center multipole expansion. The molecules are modeled as a
collection of multipole moments located at the centers of mass. Previous
calculations\cite{batista98,batista00} 
have
demonstrated that in order to reach
convergence in the multipole
expansion of the electric field at the 
relevant intermolecular distances,
the expansion had to be carried out up to and including the hexadecapole
moment. Dipole-dipole, dipole-quadrupole and
quadrupole-quadrupole polarizabilities were included to account for the
induction effects. Within this approximation, the electrostatic+induction
component takes the following form:
\begin{equation}
\fl  E_{\text{es+ind}} = - \frac{1}{2} \sum_{i}
  \left(
     \mu_{\alpha}^{i} ~ \tilde{F}_{\alpha}^{i} +
     \frac{1}{3} ~ \Theta_{\alpha\beta}^{i} ~ \tilde{F}_{\alpha\beta}^{i} 
     \frac{1}{15} ~ \Omega_{\alpha\beta\gamma}^{i} ~ \tilde{F}_{\alpha\beta\gamma}^{i} +
     \frac{1}{105} ~ \Phi_{\alpha\beta\gamma\delta}^{i} ~ \tilde{F}_{\alpha\beta\gamma\delta}^{i}
  \right) ~.
\end{equation}
Throughout this work we closely follow Stones' notation\cite{stone96}: The Einstein
convention is used for the $\alpha$, $\beta$... indices, which
run over the Cartesian components $x$, $y$ and $z$. The  $i$, $j$... indices
label the different molecules and those summations are indicated explicitly.
$\xi_{\alpha\beta\dots}^{i}$ are the static multipole moments
(see Table \ref{tab:multipoles}) defined with respect to the center of mass of
molecule $i$ and rotated along with its molecular frame.
Experimental values are used for the dipole\cite{dyke73} and
quadrupole\cite{verhoeven70} moments, while the higher moments are
obtained from \mbox{MP2/aug-cc-pVQZ} \textit{ab initio} calculations\cite{batista98}.
$\tilde{F}_{\alpha\beta\dots}^{i}$ represents the scaled electric field and
its gradients, defined by:
\begin{equation}
  \tilde{F}_{\alpha\beta\dots\nu}^{i} = \sum_{j \left( \neq i \right)}
  f_{sw}\!\left( r_{ij} \right) F_{\alpha\beta\dots\nu}^{ij}
\end{equation}
where
\begin{eqnarray}
\fl  F_{\alpha}^{ij} =
  T_{\alpha\beta}^{ij} ~ \left( \mu_{\beta}^{j} + \Delta\mu_{\beta}^{j} \right)
  - \frac{1}{3} ~ T_{\alpha\beta\gamma}^{ij} ~ \left( \Theta_{\beta\gamma}^{j} 
+ \Delta\Theta_{\beta\gamma}^{j} \right) \\ \nonumber
  + \frac{1}{15} ~ T_{\alpha\beta\gamma\delta}^{ij} ~ \Omega_{\beta\gamma\delta}^{j}
  - \frac{1}{105} ~ T_{\alpha\beta\gamma\delta\varepsilon}^{ij} ~ \Phi_{\beta\gamma\delta\varepsilon}^{j}
\end{eqnarray}
and
\begin{equation}
  F_{\alpha\beta\dots\nu}^{ij} =
    \frac{\partial}{\partial r_{\beta}} ~
    \cdots
    \frac{\partial}{\partial r_{\nu}} ~
    F_{\alpha}^{ij} ~.
\end{equation}
The interaction tensors $T$ are defined by:
\begin{equation}
  T_{\alpha\beta\dots\nu}^{ij} =
    \frac{\partial}{\partial r_{\alpha}} ~
    \frac{\partial}{\partial r_{\beta}} ~
    \cdots
    \frac{\partial}{\partial r_{\nu}} ~
    \left( \frac{1}{r} \right),~~~~~~
         \left( r \equiv r_{ij} = \vert {\bf r}_{i} - {\bf r}_{j} \vert
    \right)
\end{equation}
where $r_{ij}$ is the distance between the centers of mass of molecules $i$ and
$j$.

The induced dipole ($\Delta\mu_{\alpha}^{i}$) and quadrupole
($\Delta\Theta_{\alpha\beta}^{i}$) moments are defined by self-consistent
polarization equations:
\begin{equation}
  \Delta\mu_{\alpha}^{i} = \alpha_{\alpha\beta}^{i} ~ \tilde{F}_{\beta}^{i} +
  \frac{1}{3} A_{\alpha,\beta\gamma}^{i} ~ \tilde{F}_{\beta\gamma}^{i}
\end{equation}
\begin{equation}
  \Delta\Theta_{\alpha\beta}^{i} = A_{\gamma,\alpha\beta}^{i} ~ \tilde{F}_{\gamma}^{i} +
  C_{\gamma\delta,\alpha\beta}^{i} ~ \tilde{F}_{\gamma\delta}^{i}
\end{equation}
that are solved iteratively with a convergence threshold of $1.0\times10^{-7}$ au
for the difference between iterations for any of the components.
$ \alpha_{\alpha\beta}^{i}$, $A_{\alpha,\beta\gamma}^{i}$ and
$C_{\gamma\delta,\alpha\beta}^{i}$ are, respectively, the dipole-dipole,
dipole-quadrupole and quadrupole-quadrupole polarizabilities, shown in
Table \ref{tab:polarizabilities}. The values employed in the parametrization of our potential
were taken from the \mbox{ASP-W4} potential\cite{millot92,millot98}, i.e. the experimentally
determined\cite{gray84} values were used for the dipole-dipole polarizability,
while the dipole-quadrupole and quadrupole-quadrupole polarizabilities
were obtained from Hartree-Fock calculations and scaled by 1.25\cite{millot92}.
Since \mbox{ASP-W4} uses oxygen-centered polarizabilities and our potential locates
them in the center of mass, the values that appear in Table \ref{tab:polarizabilities}
correspond to a translational transformation of the \mbox{ASP-W4} values.

The electric field and its gradients are switched-off at short- and
long-range using the following function:
\begin{equation}
\label{eq:swf}
  f_{sw}\!\left( r \right) =
     \left\{
       \begin{array}{r@{\quad:\quad}c}
          \left[ 1 - \text{e}^{-\tau_d r} \sum_{k=0}^{6}
          \frac{\left( \tau_d r \right)^k}{k!} \right]^{1/2} &
           0 \leq r < r_{h_1} \\
          1 & r_{h_1} \leq r \leq r_{l_2} \\
          1 + x^3 \left( -6x^2 + 15x -10 \right) &
          r_{l_2} < r < r_{h_2} \\
          0                         & r_{h_2} \leq r    \\
       \end{array}
     \right.
\end{equation}
where
\begin{equation}
  x = \frac{r-r_{l_2}}{r_{h_2}-r_{l_2}} ~.
\end{equation}
The short-range damping function is used to approximately account for the penetration error
that arises from the use of a multipole expansion\cite{stone96d} at normal interaction
distances (i.e., for $r_{ij} < 5$~\AA), where the molecular charge densities are
starting to overlap significantly. A modification of the
Tang-Toennies damping function\cite{tang84} was used, where $\tau_d$ (which roughly
corresponds to the inverse decay length of the charge density in the water monomer)
was adjusted to reproduce the electric field generated by clusters and ice. It should be
noted that the application of the same damping to the electric field and its gradients
should introduce non-physical effects in the description of the interaction at
short-distances. A better approach is to redefine the interaction tensors $T$ to
include the damping\cite{millot98}, thus preserving the relation that must exist
between the electric field and its gradients. However, for the systems studied we found
that this homogeneous damping introduces only minor non-physical effects when compared
with the effects of the other approximations. Its implementation is also quite efficient. 

The long-range part of the damping function is used to make the range of the
interaction finite. Studies of the convergence of the electrostatic induction in ice
as more distant neighbors are included showed that a cutoff radius of 9 \AA\ or greater
is justifed\cite{batista98}. In order to avoid spurious forces, the potential
was switched smoothly. Based on the calculation of the induced dipole moments as
a function of the cutoff, it was found that a polynomial interpolation between
9 \AA\ and 11 \AA\ fulfilled these requirements.
\subsection{Dispersion Energy}
\label{subsec:dispcom}
The dispersion component of the interaction energy is:
\begin{equation}
  E_{\text{disp}} = - \sum_{i<j}
     \left(
       \frac{C_6}{r_{ij}^{6}} ~ g_6\!\left( r_{ij} \right) +
       \frac{C_8}{r_{ij}^{8}} ~ g_8\!\left( r_{ij} \right) +
       \frac{C_{10}}{r_{ij}^{10}} ~ g_{10}\!\left( r_{ij} \right)
     \right)
\end{equation}
where $r_{ij}$ is the O-O distance. Only the first three terms of the
dispersion expansion were included. The $C_n$ coefficients used
(Table \ref{tab:othpar}) were those recommended by Wormer and Hettema\cite{wormer1992}.
At short distance, each component is switched off by means of a Tang-Toennies
damping function\cite{tang84} similar to the one used for the electric field and gradients
(Equation~\ref{eq:swf}):
\begin{equation}
  g_n\!\left( r \right) =
  1 - \text{e}^{-\tau_d r} \sum_{k=0}^{n}
  \frac{\left( \tau_d r \right)^k}{k!} ~.
\end{equation}

\subsection{Repulsion Energy}
\label{subsec:repcom}

For the exchange repulsion, a modified Born-Mayer potential was used: 
\begin{equation}
\label{eq:rep1}
  E_{\text{rep}} = A \sum_{i<j}
                     \left(
                        1+B\!\left( \rho_i \right) + B\!\left( \rho_j \right)
                     \right)
                     r_{ij}^{-b} \text{e}^{-cr_{ij}}
\end{equation}
where $r_{ij}$ is the O-O distance and $B$ is a density-dependent term defined
by:
\begin{equation}
\label{eq:rep2}
  B \! \left( \rho_i \right) \; = \;
     \left\{
       \begin{array}{r@{\quad:\quad}c}
          0                         & \rho_i \leq 1600    \\
          \sum_{n=0}^5 a_n \rho_i^n & 1600 < \rho_i < 8000\\
          0.0875                    & 8000 \leq \rho_i    \\
       \end{array}
     \right. ~.
\end{equation}
The density of molecules at a given molecule was defined as a sum over exponential weight
functions, located at each one of the neighboring molecules:
\begin{equation}
\label{eq:rep3}
  \rho_i = C \sum_{j \left( \neq i \right)}
           \frac{\text{e}^{-d r_{ij}}}{r_{ij}^3} ~.
\end{equation}
The modification of the Born-Mayer term is purely phenomenological and arises from the
use of a single center for the exchange repulsion (i.e. the oxygen atom) instead of the
usual pure Born-Mayer terms for each atomic center. We found that the modified form used
in Equation~\ref{eq:rep1} provides a good approximation to the repulsion while 
having a simple form that is easy to implement. 
The density dependence of the repulsion was introduced to account for the changes in electron
density distribution occurring when the environment of the molecule changes from the gas phase
to condensed matter. 
As the molecule polarizes, excited electronic orbitals are partly occupied and this results in a slower decay of the electron density, thus
increasing the repulsive Pauli exchange interaction between closed shell molecules.
Such effects have, for example, been observed in atom interaction with surface adsorbates~\cite{COonPt1,COonPt2}.
The parameters used in Equations~\ref{eq:rep1}-\ref{eq:rep3}
(Table \ref{tab:othpar}) were obtained in three stages: (1) a potential energy curve
was calculated at MP2/aug-cc-pVTZ level by varying the O-O separation in the water
dimer and optimizing the structure at each point. The $B$ terms were initially
neglected and $A$, $b$ and $c$ were determined by fitting Equations \ref{eq:rep1}-\ref{eq:rep3}
to the difference between the MP2 potential energy curve and the sum of the electrostatic
and dispersion contributions previously described. The parameters were constrained to
give the same minimum as the MP2 curve used for the fitting.
(2) The $B$ terms were then introduced and the limit value
($\rho_i \geq 8000$) was varied to obtain the correct cohesive energy and
cell parameters for ice Ih (see \ref{subsec:ice}).
(3) Finally, a polynomial interpolation was introduced between $0$ and the
limit value in order to provide balanced results for a few clusters of
intermediate densities.
This polynomial was adjusted to obtain the best possible binding energy
and structure for the (H$_2$O)$_{n}$ with $n=3$ to $6$ ring clusters
(see \ref{subsec:wnres}). The parameter $d$ used in the density of molecules
( Equation \ref{eq:rep3})
was chosen so as not to introduce a large distinction between clusters, surface molecules and
bulk molecules. This decay length yields a density whose main contribution is associated with
the nearest-neighbor molecules (at distances between 2.7 and 3.0\AA), while the contribution
from the next-nearest-neighbors is 8\% of that provided by the nearest-neighbor. The 
more distant molecules only give a minor contribution to this term.


\section{Results and Discussion}
\label{sec:resul}

\subsection{The Water Dimer}
\label{subsec:w2pes}

\begin{figure*}
\includegraphics[scale=0.20]{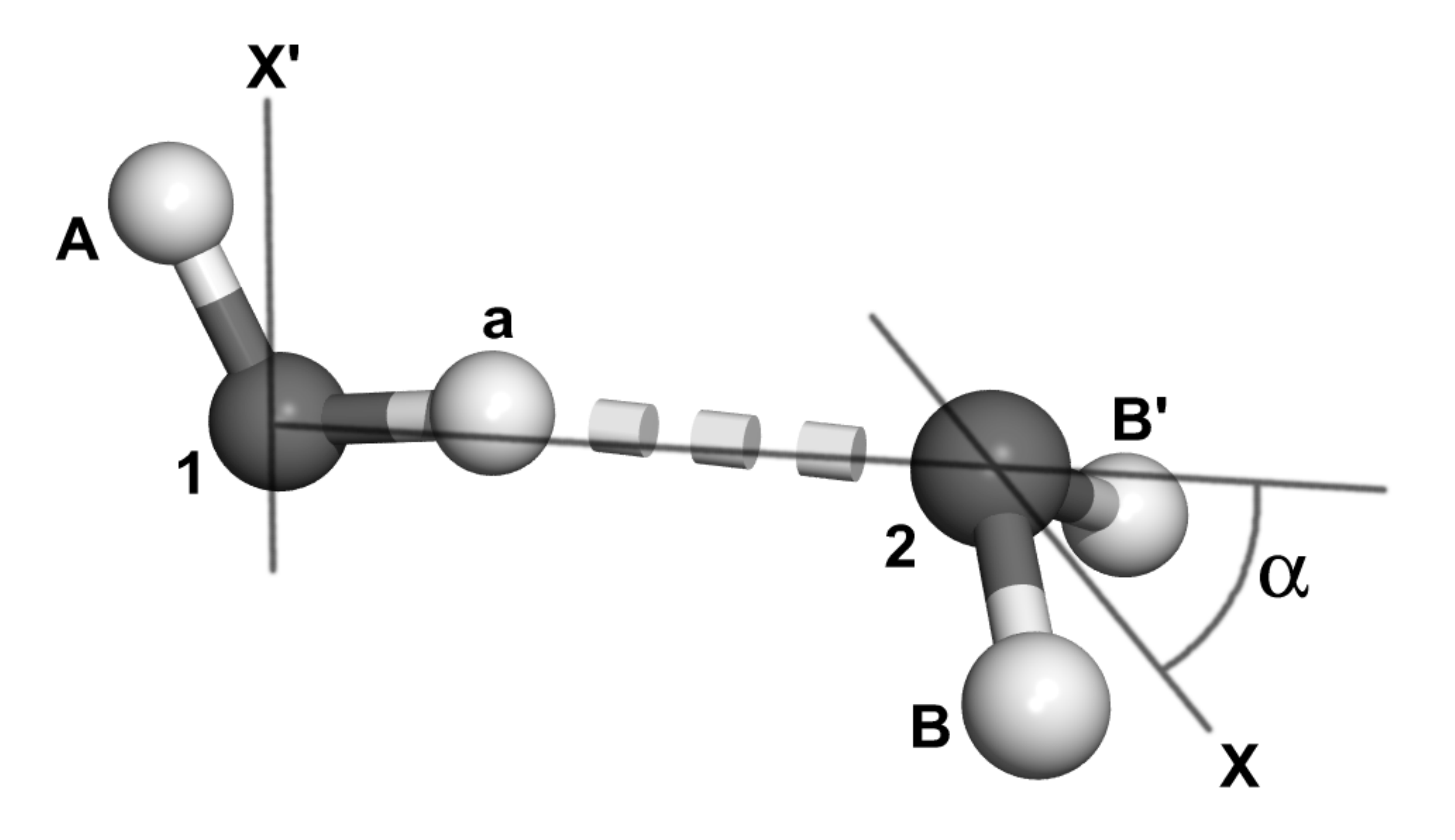}
\caption{\label{fig:w2geom}
         Water dimer in its optimal configuration. See Table \ref{tab:w2geom}
         for structure details.
        }
\end{figure*}

A close analysis of the structure and potential energy curves (PECs) of the
water dimer is of special interest since most anomalies in the interaction
potential would be easiest to recognize in this simple system. Figure
\ref{fig:w2geom} shows the water dimer in its optimal configuration while Table
\ref{tab:w2geom} presents a comparison between the results predicted by our
potential, the NCC\cite{niesar90} and \mbox{ASP-W4}\cite{millot92,millot98} potentials,
and \textit{ab initio} MP2/aug-cc-pVTZ
results. The \mbox{ASP-W4} and NCC calculations were performed using
Orient 3.2\cite{stone97}, while the Gaussian 98\cite{frisch98} package was used for the
\textit{ab initio} calculations. SCME and \mbox{ASP-W4} give rather similar results.
The main errors observed for the latter are the 0.06~\AA\ overestimation of
the r$_{OO}$ distance (a problem that is also found on the larger clusters)
and the buckling of the hydrogen bond in the wrong direction. The NCC potential also shows
an overestimation of the O-O distance and a rather large overestimation
of the wagging angle of the acceptor monomer (1,2,X). Finally, the largest error shown by
SCME ocurrs for the (1,2,X) angle which is underestimated by about 9\deg.

\begin{figure}
\includegraphics[scale=0.50]{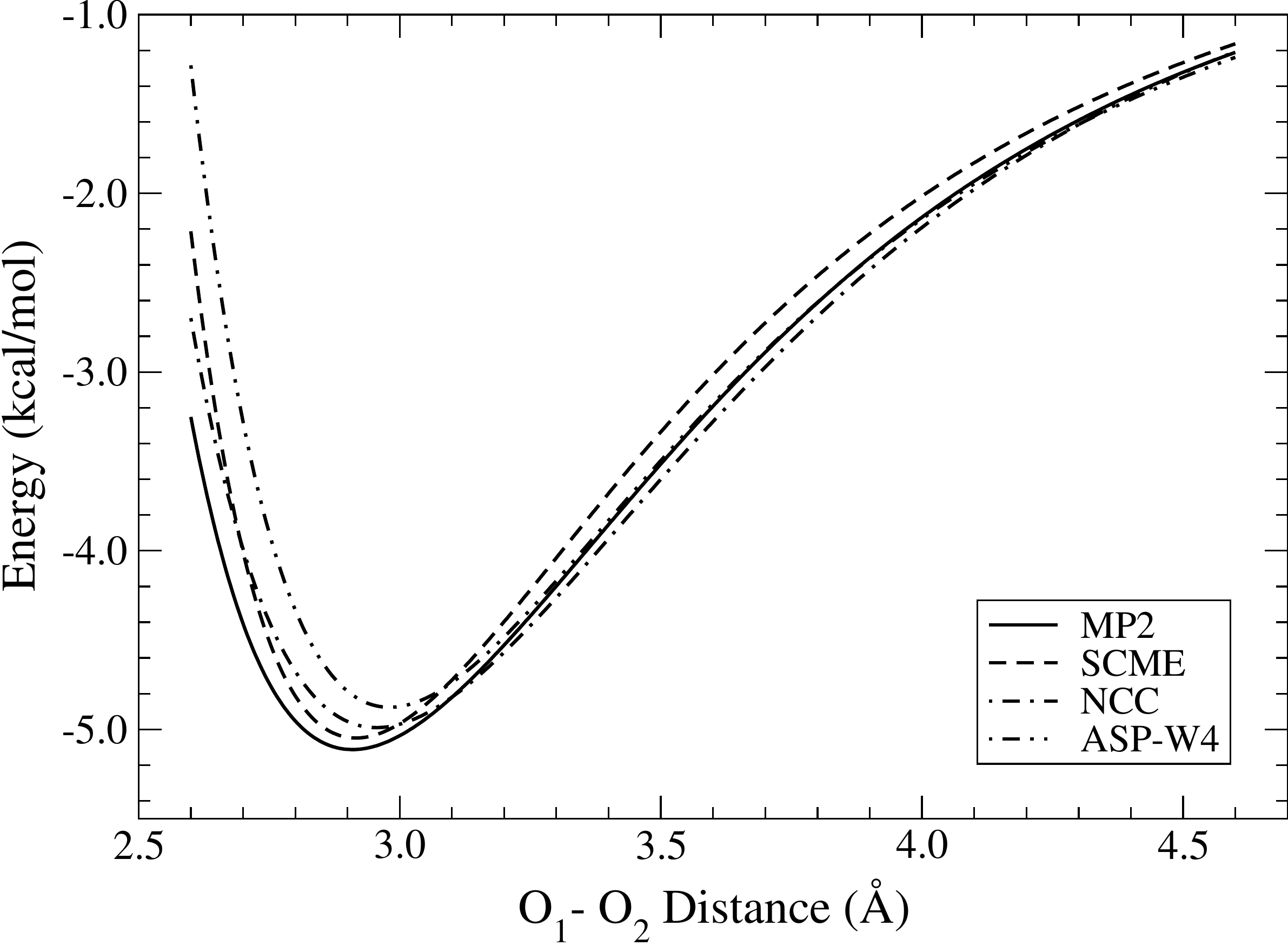}
\caption{\label{fig:croo}
         Comparison of the potential energy curves for (H$_2$O)$_2$ calculated
         with our model potential and several other methods. The O-O
         distance was varied while the rest of the structure was kept at
         its optimal configuration. See Figure \ref{fig:w2geom} for
         structure details.
}
\end{figure}

\begin{figure}
\includegraphics[scale=0.50]{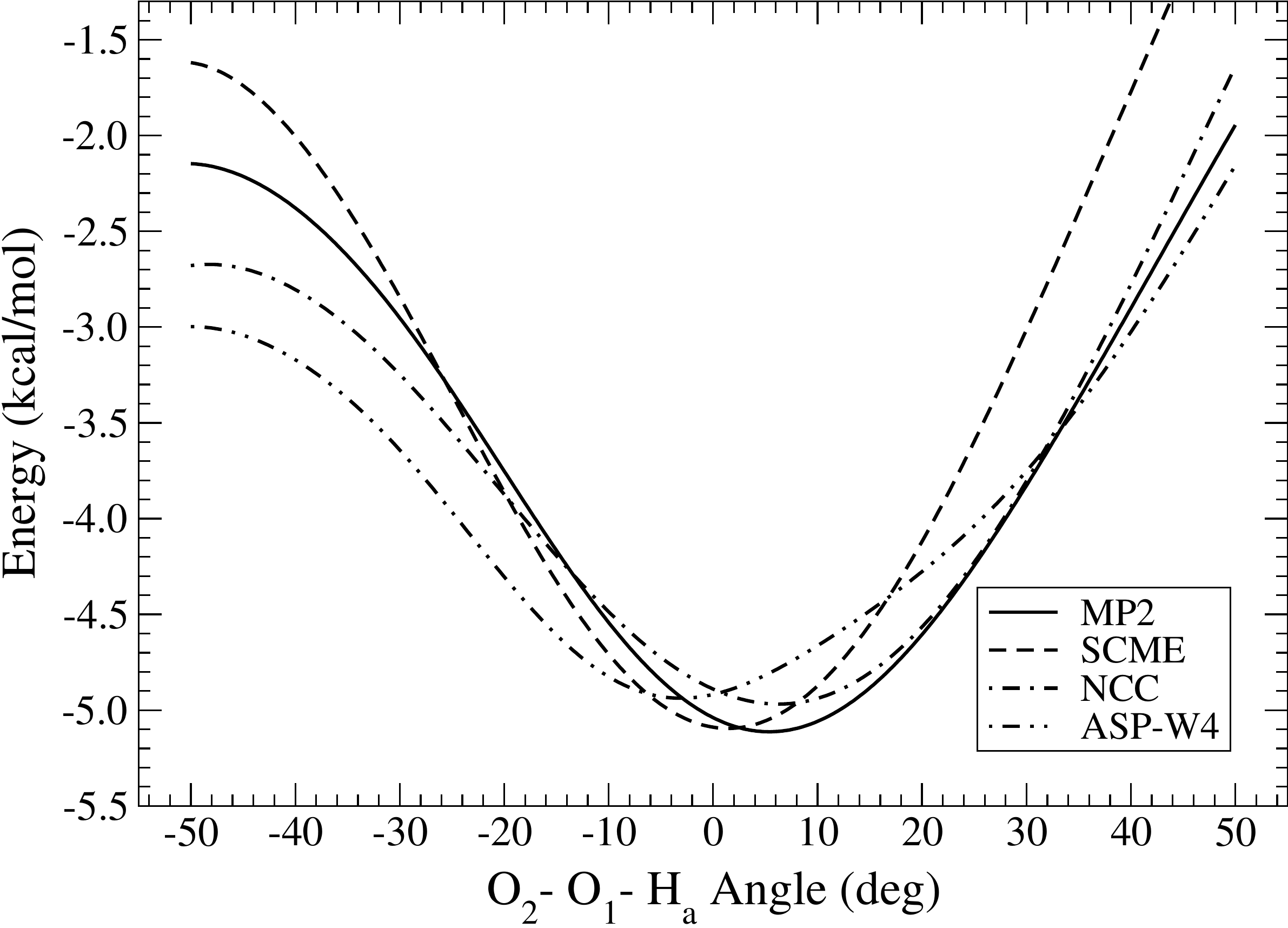}
\caption{\label{fig:caooh}
         Comparison of the potential energy curves for (H$_2$O)$_2$ calculated
         with our model potential and several other methods. The hydrogen
         bond angle was varied while the rest of the structure was kept at
         its optimal configuration. See Figure \ref{fig:w2geom} for
         structure details.
}
\end{figure}

\begin{figure}
\includegraphics[scale=0.50]{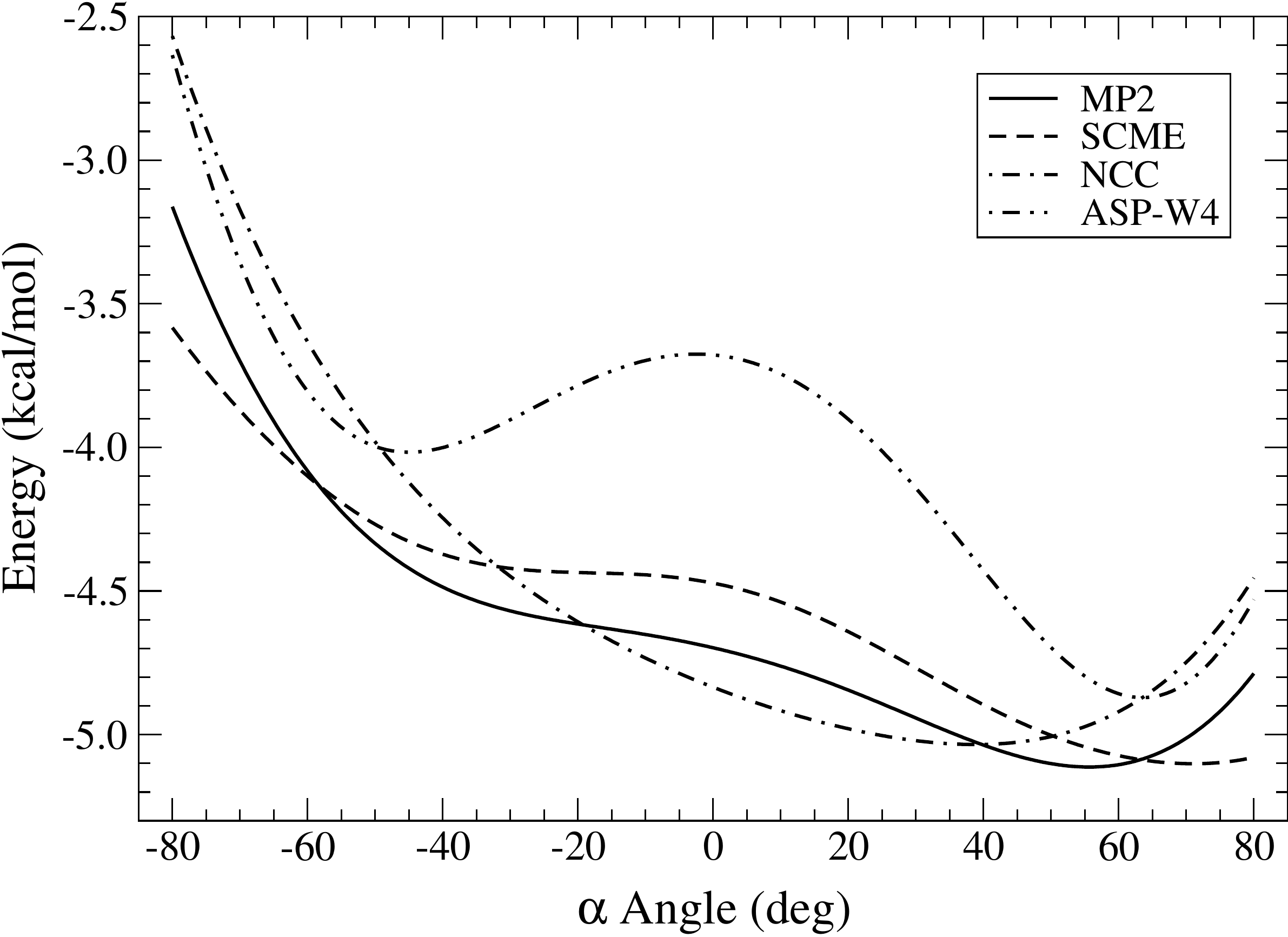}
\caption{\label{fig:caoox}
         Comparison of the potential energy curves for (H$_2$O)$_2$ calculated
         with our model potential and several other methods. The acceptor
         monomer wagging angle was varied while the rest of the structure
         was kept at its optimal configuration. See Figure \ref{fig:w2geom}
         for structure details.
}
\end{figure}

\begin{figure}
\includegraphics[scale=0.50]{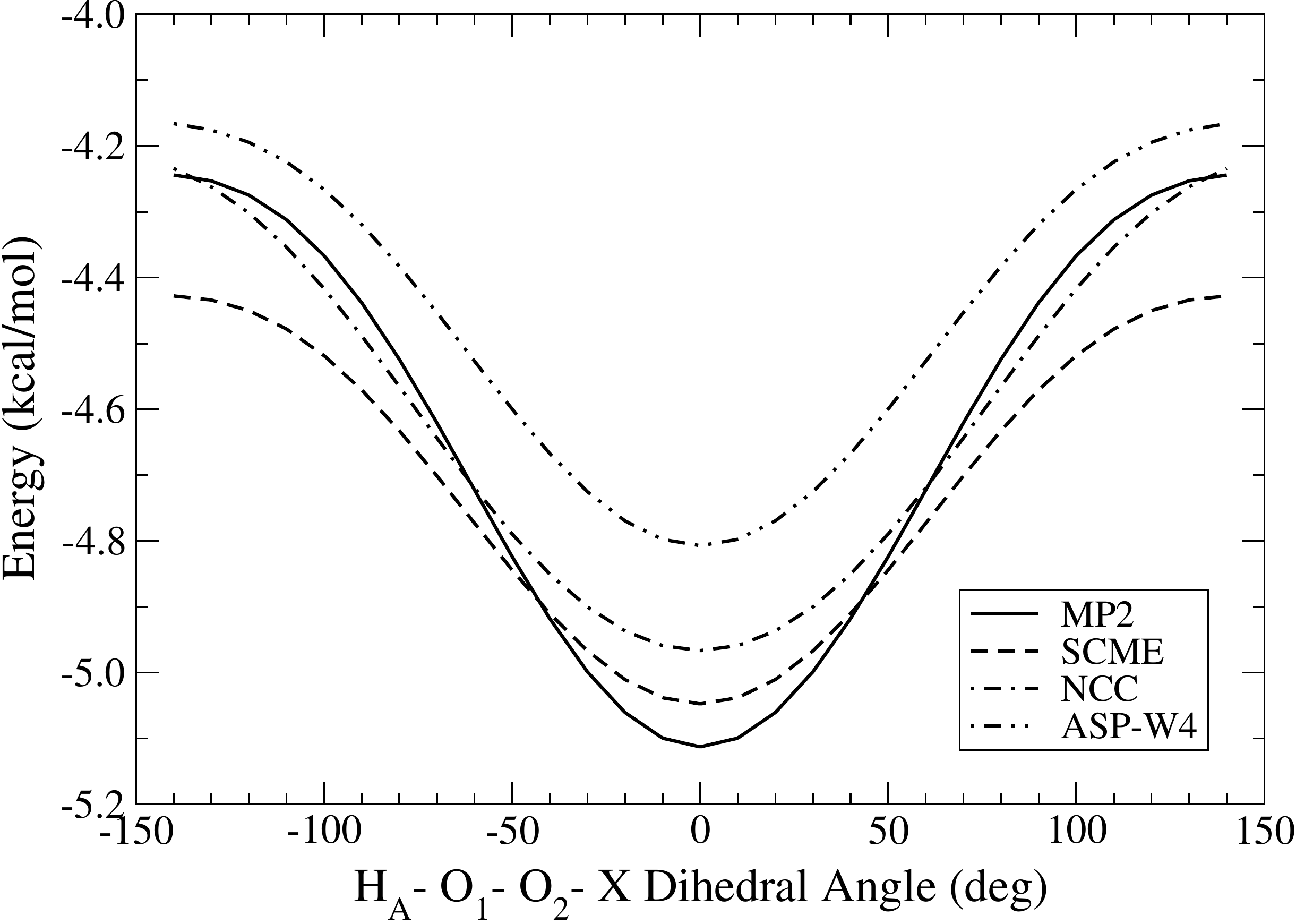}
\caption{\label{fig:cdhohx}
         Comparison of potential energy curves for (H$_2$O)$_2$ calculated
         with our model potential and several other methods. The free
         hydrogen in the donor monomer was rotated around the hydrogen bond
         while the rest of the structure was kept at its optimal
         configuration. See Figure \ref{fig:w2geom} for structure details.
        }
\end{figure}

\begin{figure}
\includegraphics[scale=0.50]{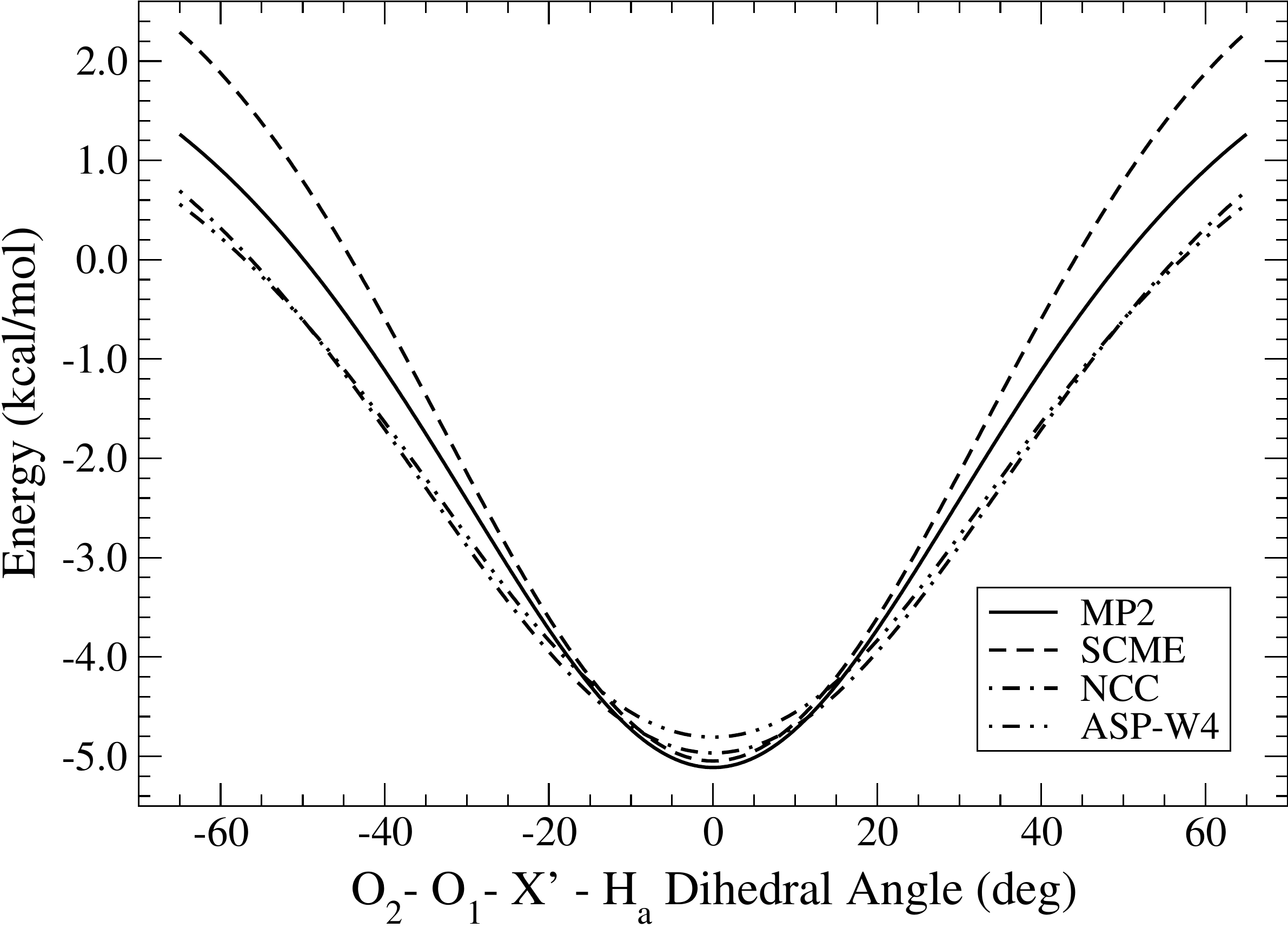}
\caption{\label{fig:cdooxh}
         Comparison of potential energy curves for (H$_2$O)$_2$ calculated
         with our model potential and several other methods. The acceptor
         monomer was rotated around the donor monomer
         while the rest of the structure was kept at its optimal
         configuration. See Figure \ref{fig:w2geom} for structure details.
        }
\end{figure}

Figures \ref{fig:croo}-\ref{fig:cdooxh} show the PECs for the
deformation of the water dimer along five coordinates of special interest. These
curves were obtained by varying a given coordinate while keeping the rest
of the structure fixed at the optimal MP2 values.
Figure \ref{fig:croo} shows that, in the long-range regions (r$_{OO} >$ 3.2 \AA\ ), these
potentials are essentially equivalent, a consequence of the similarity between the
electrostatic+induction components used by each of them. Some differences appear for the
short-range interaction region, although in general they are well within the expected
accuracy of the models. The most important exceptions to this observation occur for
the variation of the hydrogen bond angle and the acceptor monomer wagging angles
(Figures \ref{fig:caooh} and \ref{fig:caoox}).

In the first case (Figure \ref{fig:caooh}), the NCC and SCME potentials
behave similarly in the minimum region, with the NCC potential showing
the best overall agreement with the \textit{ab initio} results. The
deviation shown by \mbox{ASP-W4} is small but significant since the buckling of
the hydrogen bond is in the opposite direction to that predicted by the
\textit{ab initio} calculations. For larger deformations of
the angle, however, the potentials show some notorious differences. For
example, the predicted barrier for the switching of the hydrogen bond from
H$_{\text{a}}$ to H$_{\text{A}}$ varies by 1.5 kcal/mol. This is especially
true for \mbox{ASP-W4}, which underestimates the barrier by almost
1 kcal/mol. For the deformation in the opposite direction, the largest
deviation is observed for the SCME potential, which overestimates the repulsion
between the lone electron pairs of each monomer.

Figure \ref{fig:caoox} shows the variation of the acceptor monomer wagging
angle. For this distortion, the best behavior is observed for our new
potential, which correctly reproduces the shoulder corresponding to the
transference of the hydrogen bond from one lone electron pair to the other.
In the case of the NCC potential this shoulder is completely missing,
predicting an equilibrium angle that is smaller than the one
predicted by the MP2 results. The most striking feature is the barrier
predicted by the \mbox{ASP-W4} potential. This barrier and the minimum
observed around -45\deg\ are the result of the restriction of the O-O distance
to a value smaller than the optimal for \mbox{ASP-W4}. When the PEC
is calculated at a longer distance, these features disappear.

Since the only non-spherically symmetric contributions to the SCME potential energy
arise from the electrostatic+induction terms, the differences
observed between the \textit{ab initio} and SCME results must be associated
with these terms. Moreover, since the multipole moments used in the model
potential are essentially identical to those obtained at MP2 level, we
conclude that the origin of the differences must lie in either the damping
function used for the electric field and its gradients, or in the quality of the
induction approximation used. 


\subsection{The (H$_2$O)$_{n}$ Clusters with $n=$3 to 6}
\label{subsec:wnstruct}

\subsubsection{Ring Clusters}
\label{subsec:wnres}

\begin{figure}
\includegraphics[scale=0.20]{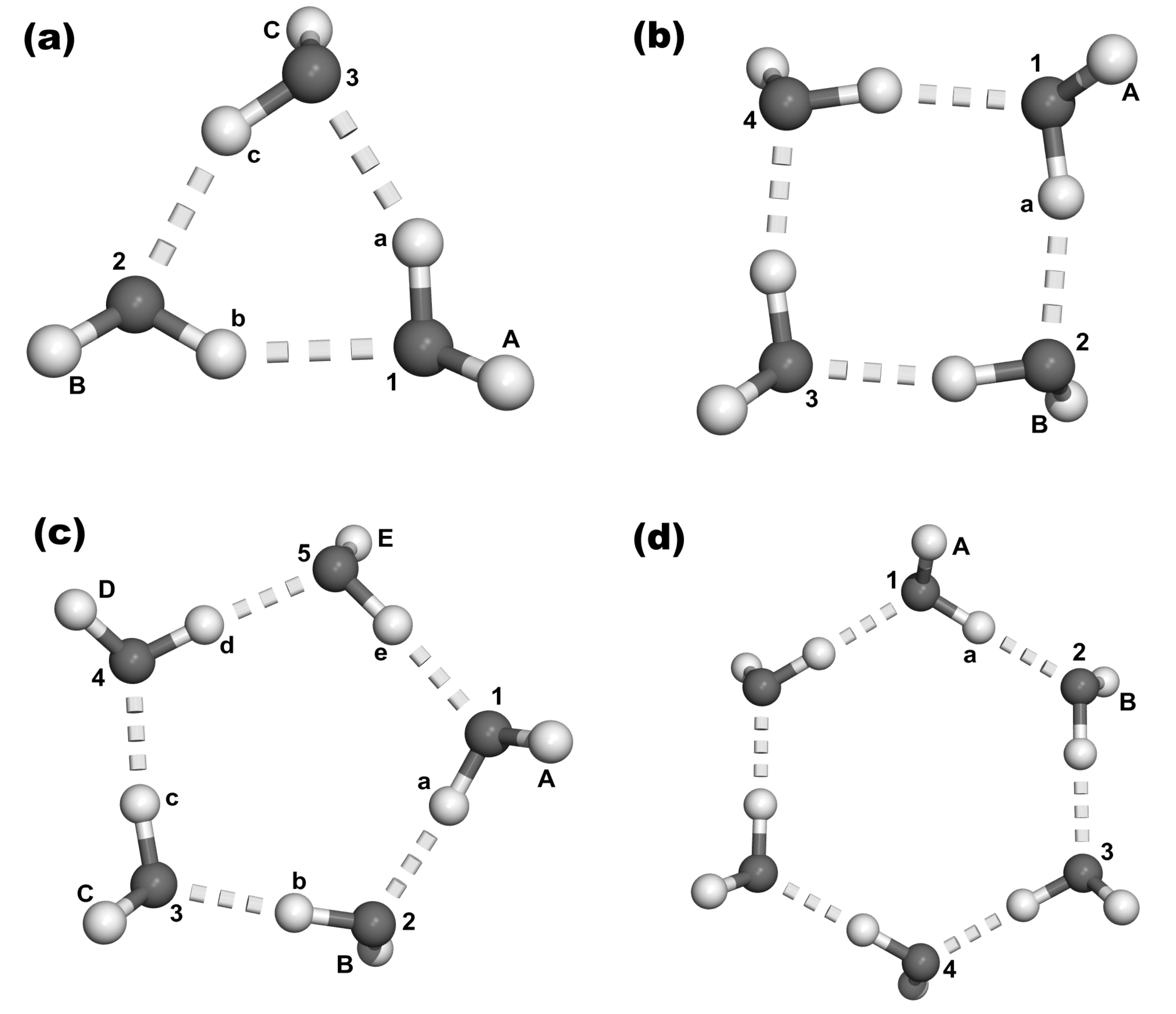}
\caption{\label{fig:w3-6geom}
         Optimal conformations of the (H$_2$O)$_{n=3-6}$ clusters. See Tables
         \ref{tab:w3geom}-\ref{tab:w6geom} for structure details.
        }
\end{figure}

The next important test of the potential function
comes from the comparison of
the predicted structures of the small ring clusters to those obtained with
\textit{ab initio} methods. Tables \ref{tab:w3geom}-\ref{tab:w6geom} present
these results and Figure \ref{fig:w3-6geom} explains the labels
used. We have divided the analysis of the results into different types of
coordinates, i.e., O-O distances, hydrogen-bond angles (1,a,2), O-framework
dihedrals (1,2,3,4), free hydrogen-O-framework dihedrals (A,1,2,3) and
free hydrogen-free hydrogen dihedrals (A,1,2,B).

\begin{figure}
\includegraphics[scale=0.50]{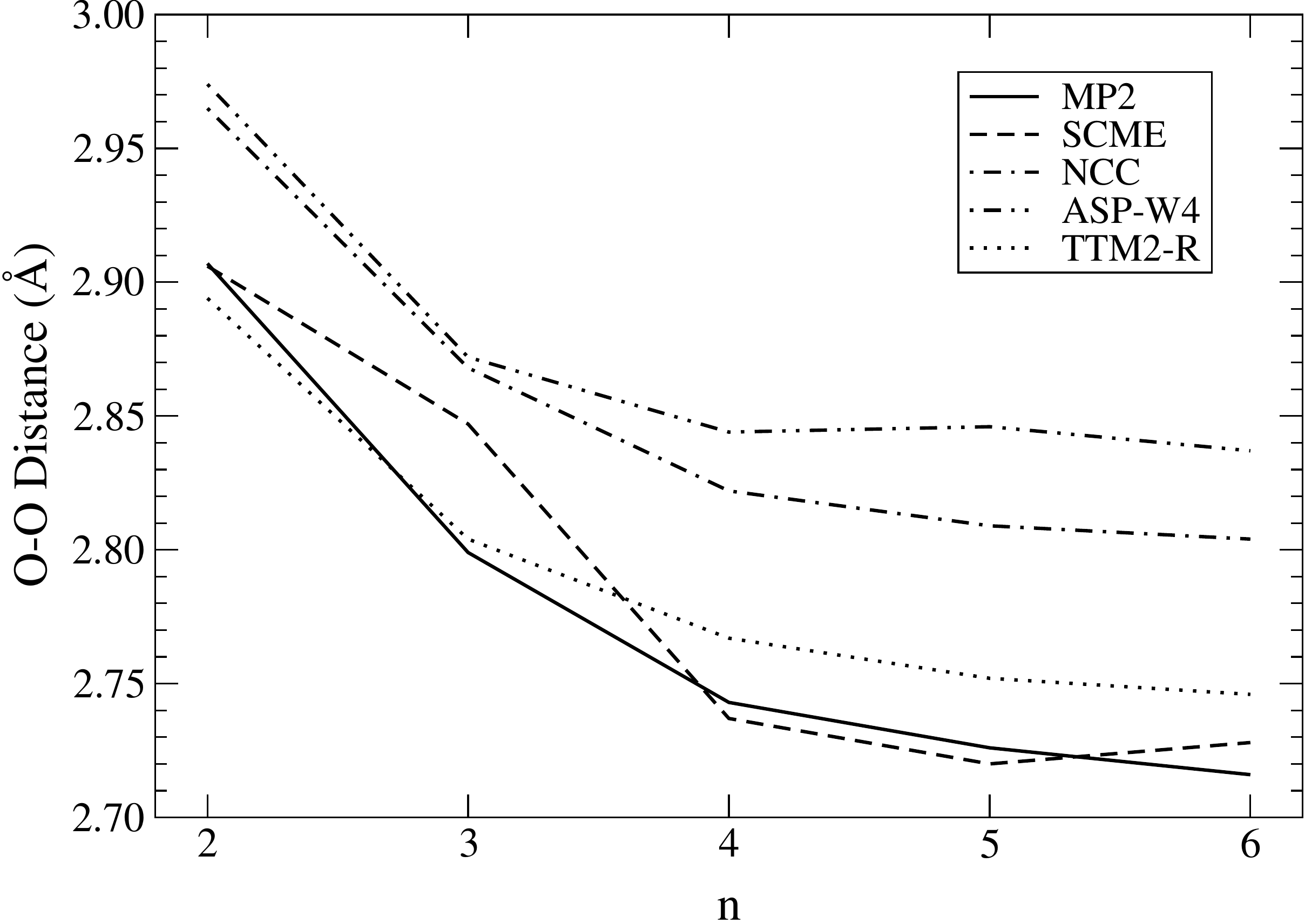}
\caption{\label{fig:wndoo}
         Comparison of the average O-O distance for the
         (H$_2$O)$_{n=2-6}$ clusters calculated with our model potential
         and several other methods.
        }
\end{figure}

The SCME potential gives good results for the O-O distance, with a 
mean absolute deviation with respect to the MP2 values of only 0.02 \AA. This
is to be compared with the  0.08 and 0.10 \AA\ deviations shown by NCC and
\mbox{ASP-W4}, respectively. This good agreement 
can be seen
in Figure \ref{fig:wndoo}, which compares the average O-O distance for each
cluster calculated with the methods mentioned above. It is clear that, with the
exception of the trimer, the SCME potential provides 
an accurate description of the variation in the
O-O distance, while NCC and \mbox{ASP-W4} largely give an overestimate.

Angles between hydrogen-bonds are best described
by the NCC potential with a mean absolute deviation of 2\deg. The SCME and
\mbox{ASP-W4} potentials give slightly larger deviations of 5\deg\ and 6\deg,
respectively. Perhaps the most striking result is the rather large
error (about 11\deg) shown by the \mbox{ASP-W4} potential for the water hexamer.
The SCME potential provides good estimates for the three
different types of dihedral angles studied. In the case of the O-framework
dihedral angles we obtain significantly lower deviations than those obtained
with the other potentials. This is especially true in the case of NCC, which
shows large errors in those dihedrals for all clusters. Moreover, although both
\mbox{ASP-W4} and NCC show similar mean deviations, the former shows a very
large (about 25\deg) error in the case of the hexamer. For the other dihedrals
our results are similar to those obtained with \mbox{ASP-W4} and significantly
better than the NCC results.

\begin{figure}
\includegraphics[scale=0.50]{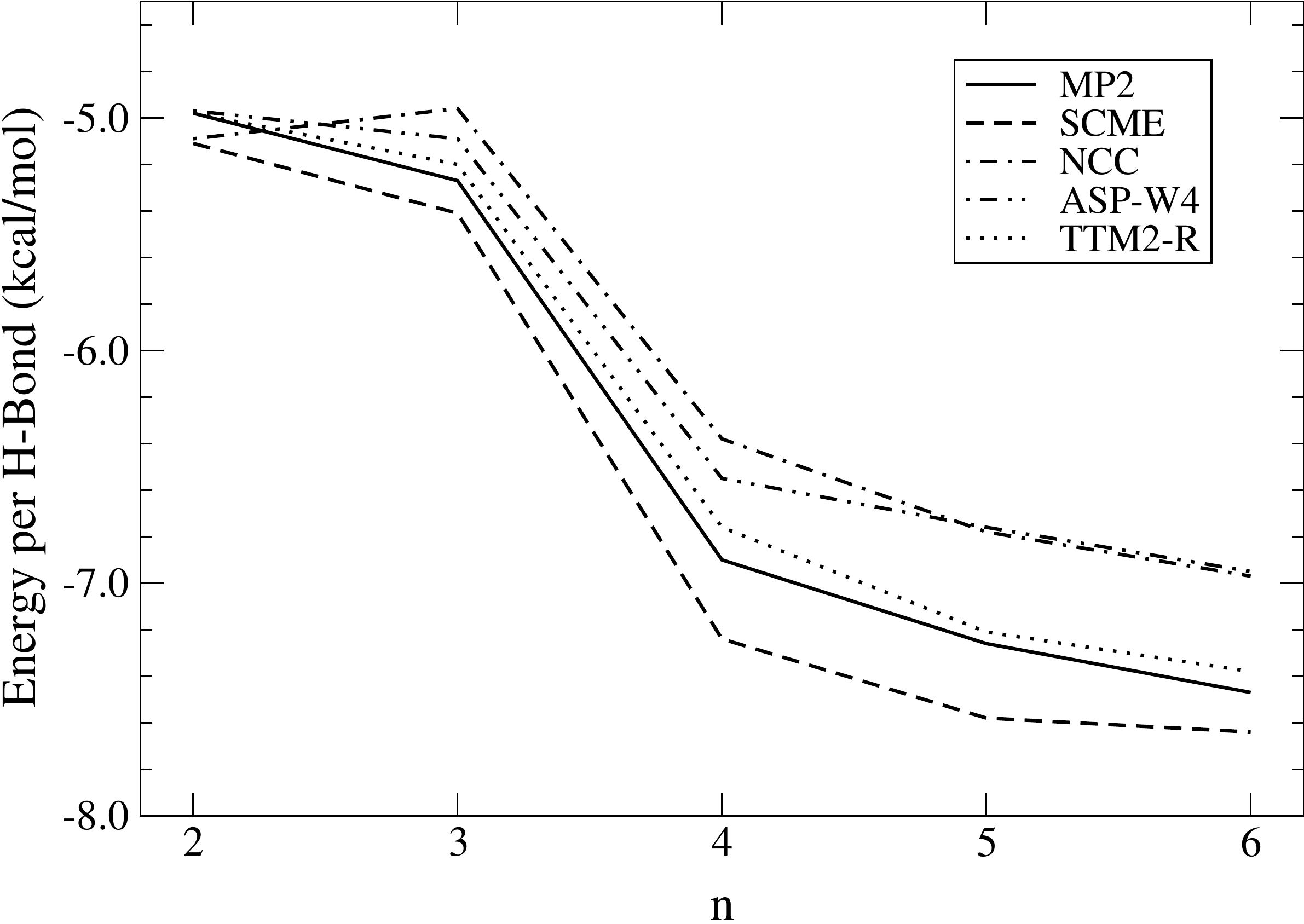}
\caption{\label{fig:wnephb}
         Comparison of the interaction energy per hydrogen bond for the
         (H$_2$O)$_{n=2-6}$ clusters calculated with our model potential
         and several other methods. MP2 and TTM2-R energies taken from Refs.
         \cite{burnham02c} and \cite{burnham02a}, respectively.
        }
\end{figure}

In Table \ref{tab:wnenergies} we present the interaction energy for the
(H$_2$O)$_{n=3-6}$ clusters at their optimized geometry. Also included for comparison
are results for the water dimer. As in the previous section,
we compare our results to those obtained with the ASP-W4 and NCC model potentials.
We also include \textit{ab initio}
MP2/CBS results\cite{burnham02c} and TTM2-R\cite{burnham02a,burnham02b} results taken
from the literature. The mean absolute deviation
between our potential and the MP2 results is only 0.9 kcal/mol, about half
of the deviation observed for both NCC and \mbox{ASP-W4} (1.6 kcal/mol). It is,
however, larger than the one observed for the TTM2-R potential (0.3 kcal/mol)
\cite{burnham02a}.
Figure \ref{fig:wnephb} shows the variation of the interaction energy per
hydrogen bond with the size of the cluster. Both NCC and \mbox{ASP-W4}
underestimate the interaction energy, with this underestimation increasing for
the larger clusters, while TTM2-R does an excellent job in predicting the
interaction energies of these clusters. Our new potential consistently
overestimates the interaction energy by about 0.2 kcal/mol per hydrogen bond.
This deviation, probably related to the functional form used for the repulsion
component, is also found for the energies of some isomers of the water
hexamer (see Section \ref{subsubsec:w6isom}).

For all the structural parameters described above, the largest differences between
the SCME and
MP2 results occur for the water trimer, a cluster that poses a special challenge
for our potential. For example, if only (H$_{2}$O)$_{n=4-6}$ are considered,
the deviation of the SCME O-O distances from the \textit{ab initio} results
is only 0.008 \AA. Similarly, the free hydrogen-free hydrogen dihedral angles
show a mean deviation of almost 15\deg\ for the trimer, but are only 6\deg\ for
the hexamer.
These discrepancies arise from a mixture of problems 
that we believe originate from the
repulsion component. First, (H$_{2}$O)$_{3}$
has a strained structure that takes the potential into regions where the
dimer-based parametrization of the two-body part of the repulsion is less accurate.
Second, for the parametrization of
the density-dependent repulsion term, we assumed that the $B$ parameter increases
monotonically with the local density at each of the 
monomers. This approximation
is not adequate in the case of the ring structure of the trimer.
There is a subtle balance
between the O-O distances and the dihedral angles that is controlled by the strength
of the repulsion between the monomers.  


\subsubsection{The Cage, Prism, Book and Ring Isomers of (H$_2$O)$_6$}
\label{subsubsec:w6isom}

\begin{figure}
\includegraphics[scale=0.20]{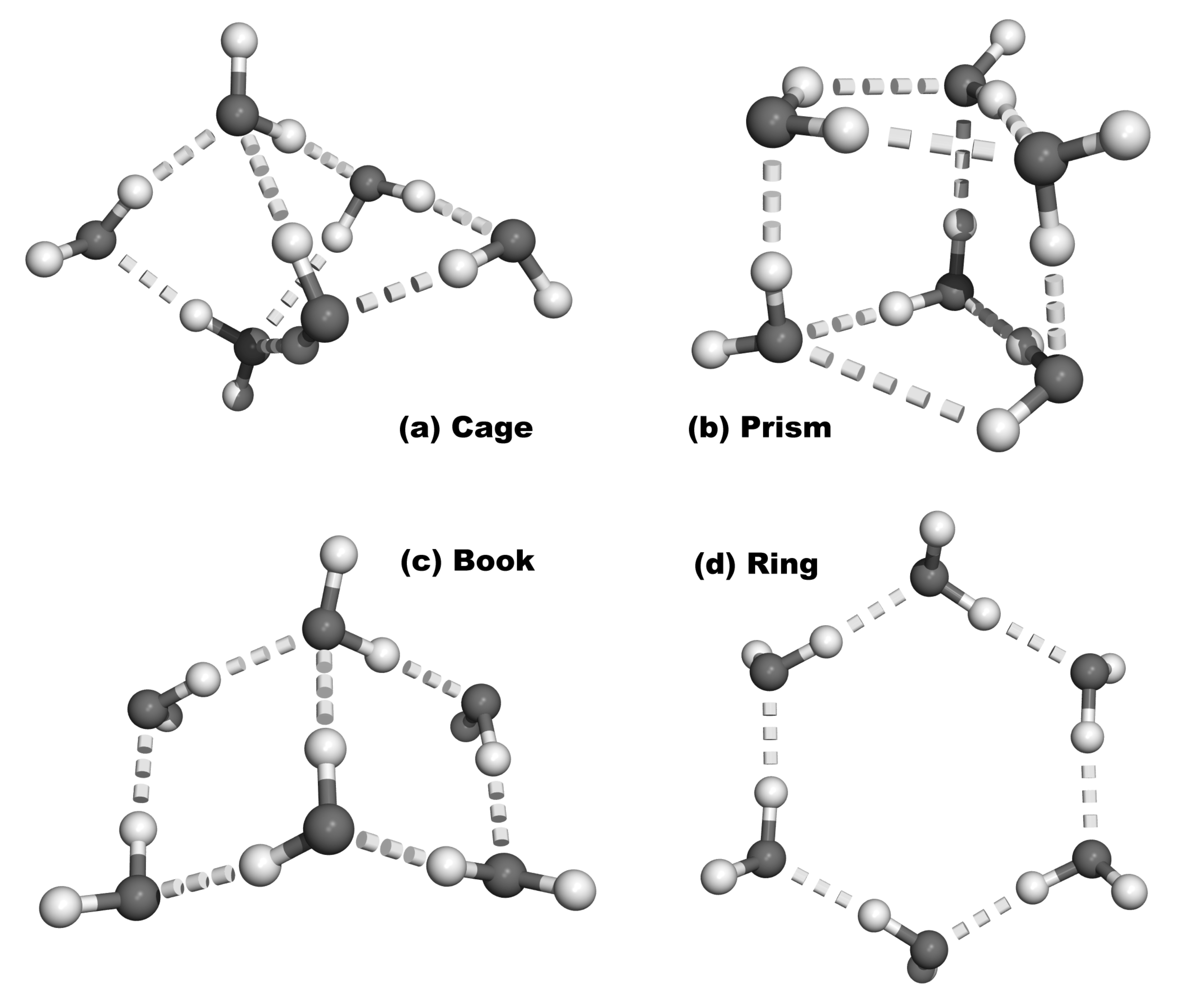}
\caption{\label{fig:w6oth}
         Structure of the cage, prism, book and ring isomers of (H$_2$O)$_6$.
        }
\end{figure}

Of all the stable conformations of (H$_2$O)$_6$, the so-called prism, cage, book and ring
isomers (Figure \ref{fig:w6oth}) have become a benchmark to test new water potentials. The
near degeneracy and difference in structure make them ideal to discover imbalances and
problems in model interaction potentials. Table \ref{tab:w6othenerg} presents a comparison of
the theoretical interaction energies of these clusters calculated with the SCME potential to
the same methods discussed in the previous section, and in addition to more recent $\Delta$CCSD(T)
results computed by adding the MP2-CCSD(T) energy difference at the triple-zeta basis set level 
to the complete basis set (CBS) MP2 results\cite{klimes10}.
The \mbox{TTM2-R} potential provides the
best results with a mean absolute deviation of 0.5 kcal/mol. The \mbox{ASP-W4} and SCME
potentials have similar deviations ($\sim$1.6 kcal/mol), while for NCC the results are
slightly less accurate (deviation $\sim$2.0 kcal/mol). The ring isomer is predicted as the
least stable of all the structures by all the potentials used, in agreement with the $\Delta$CCSD(T)
results. The relative stability of the remaining isomers is less clear due to their very
similar energies. Although the \mbox{ASP-W4} and NCC potentials give the correct energetic
ordering for the different isomers (i.e. $ E_{prism} < E_{cage} < E_{book} < E_{ring}$), the
book isomer is predicted to be too loosely bound relative to the prism and cage isomers when
comparing with the $\Delta$CCSD(T) results.
On the other hand, both the \mbox{TTM2-R} and SCME potentials give a dispersion of the energies 
in better agreement with 
the \textit{ab initio} results. The errors observed for the
SCME potential are consistent with the systematic overestimation of the binding energies
discussed in the previous section. For each of the hexamer isomers, the error is $\sim$0.2
kcal/mol per hydrogen bond. If this systematic error is removed by applying a constant
correction to each bond energy, the mean absolute deviation of the total energies predicted
with our potential is only 0.4 kcal/mol.


\subsection{Liquid Water}
\label{subsec:lwater}

The SCME potential is intended to be applicable over a wide range of
configurations of the molecules including those of liquid water, even though 
no information about liquid water was used in the development of the potential function. 
The properties of liquid water calculated with the
SCME potential function therefore represent a prediction. 
Canonical molecular dynamics simulations were carried out at 298K 
using a cubic cell of 19.72 \AA\ per side, containing 256 molecules. 
Four uncorrelated initial configurations were extracted from a previous classical 
force field simulation.
The step used in the integration of the equations of motion was
2 fs. Each cell was equilibrated until the average of the total energy was 
observed to remain constant, after which statistics were 
collected for 400 ps.
During the equilibration period, the temperature was reset to 298K every
50 fs by redistributing the translational and rotational velocities of all
the molecules according to a Boltzmann distribution~\cite{Andersen80}. 
During the collection
period, the temperature was kept constant at 298K by readjusting the velocity
of single molecules every 50 fs.
The computational time needed for a simulation of 500 time steps on a single core Intel Xeon 3.5 GHz 
processor was 20 min. 

\begin{figure}
\includegraphics[scale=1.0]{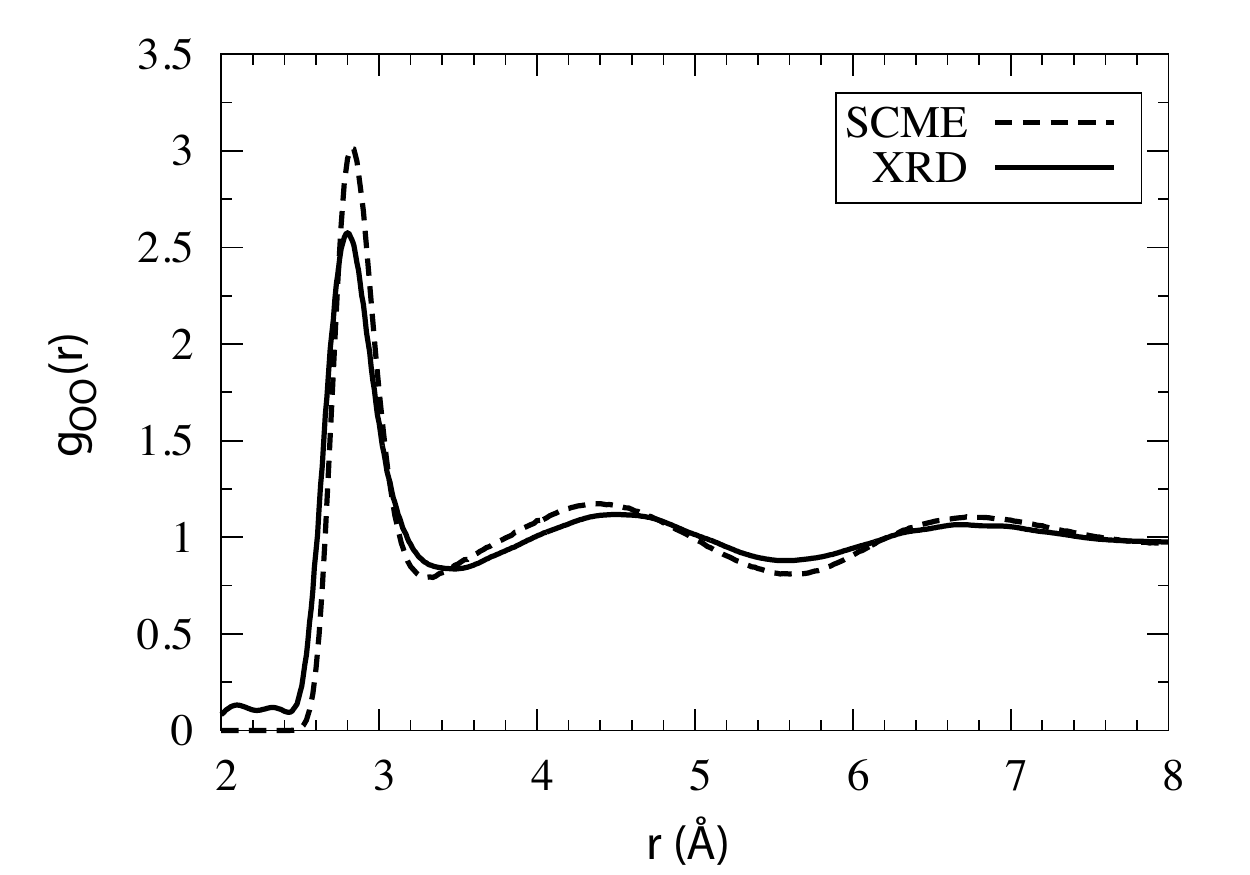}
\caption{\label{fig:goo}
         Comparison of experimental\cite{skinner13} and theoretical O-O radial distribution
         functions of liquid water at 298K.
        }
\end{figure}

\begin{figure}
\includegraphics[scale=1.0]{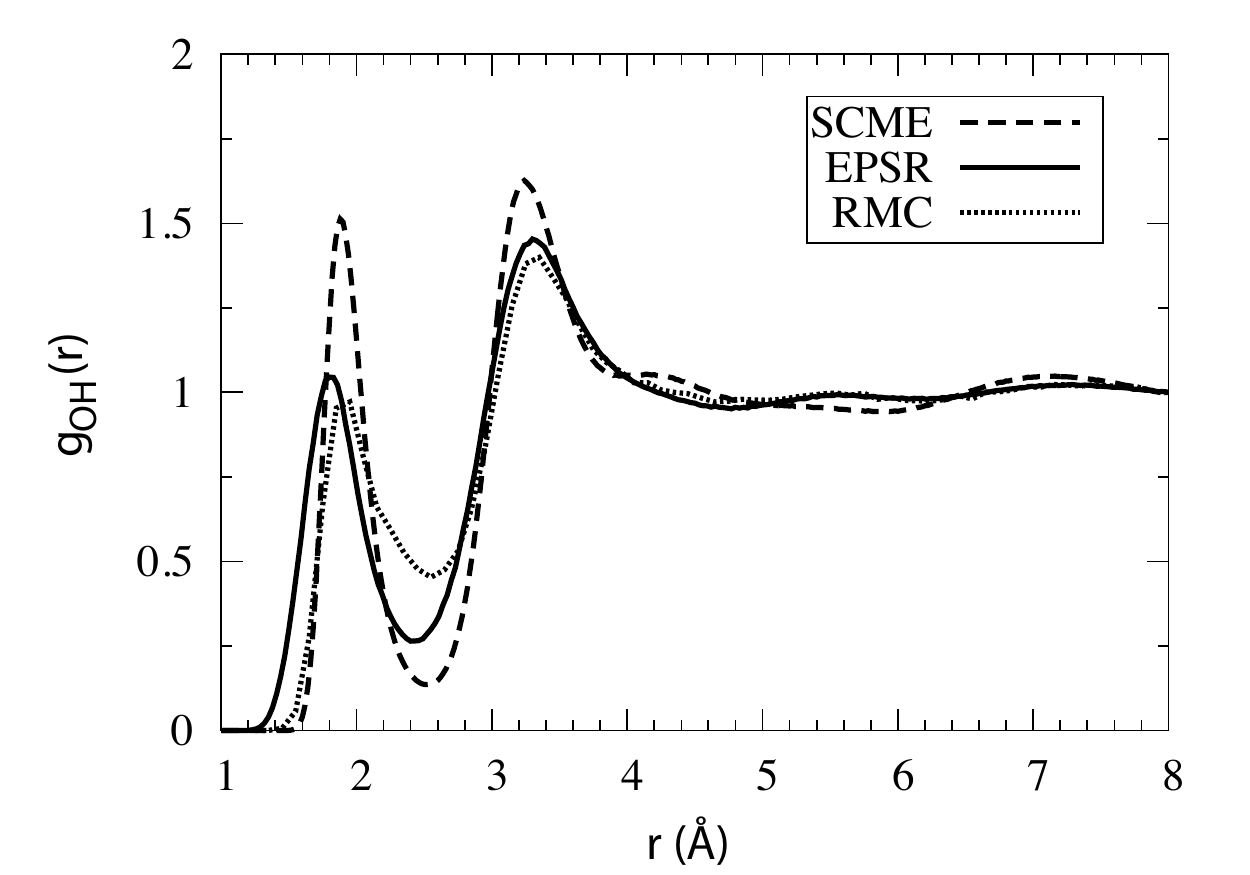}
\caption{\label{fig:goh}
         Comparison of experimental\cite{soper07,wikfeldt09} and theoretical O-H radial distribution
         functions of liquid water at 298K.
        }
\end{figure}

\begin{figure}
\includegraphics[scale=1.0]{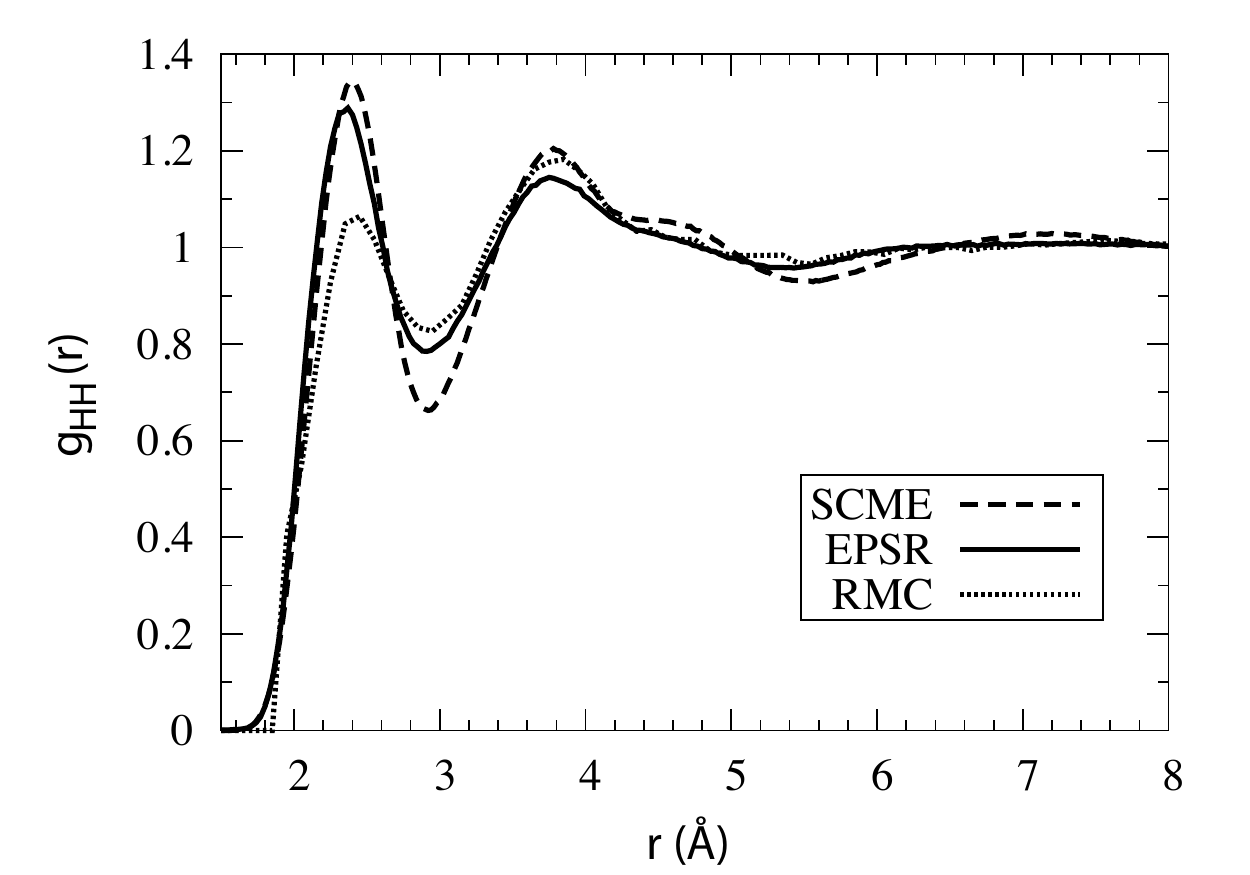}
\caption{\label{fig:ghh}
         Comparison of experimental\cite{soper07,wikfeldt09} and theoretical H-H radial distribution
         functions of liquid water at 298K.
        }
\end{figure}

From the simulated trajectory, we generated the
radial distribution functions (RDFs). Each run resulted in
very similar distribution functions, thus confirming the independence of the
final result from the initial configuration.
Figures \ref{fig:goo}-\ref{fig:ghh} show the O-O, O-H and H-H 
RDF curves, obtained by averaging the four runs performed.
An O-O curve obtained from a systematic study of x-ray diffraction 
datasets\cite{skinner13} is also shown as well as 
O-H and H-H curves obtained from EPSR\cite{soper07} and RMC\cite{wikfeldt09}
structure refinement of x-ray and neutron scattering experiments.
The agreement between experiment and our theoretical results is rather good
for each of the three RDF curves, especially in view of the fact that our 
potential function has in no way been adjusted to reproduce such data and 
considering that the experimental RDF curves contain uncertainties. 
Two main differences between our simulation and the experiments
are a shift of the second peak in the O-O curve to shorter distances
and more structured long-range regions predicted by our potential. For better
comparison we should carry out quantum mechanical simulations rather than 
classical simulations since a significant softening of the structure 
may occur\cite{kuharski84,lobaugh97,morrone08}.
Indeed, in a recent series of path-integral simulations\cite{morrone08} it was found
that the first peak of the O-O g(r) was lowered by about 0.4 compared to classical dynamics
simulation, which corresponds closely with the discrepancy in peak height observed
here in Figure~\ref{fig:goo}.

The definition of the electric properties of a molecule embedded in a condensed phase
is subject to ambiguity. The difficulty of arriving at meaningful values
for these quantities by use of \textit{ab initio} methods has been pointed
out\cite{batista99a}. 
A recent theoretical estimate for ice gave significantly larger values
than previous estimates, ca. 3.1 Debye\cite{batista98}.
Our calculations of liquid water with the SCME potential give an average molecular moment of
2.96 $\pm$ 0.26 Debye (obtained by averaging over all the cells
used and over all the molecules in each cell). This value is in good agreement with a
density functional theory estimate of 2.95 Debye\cite{silvestrelli99}.
The TTM2-R model potential, on the other hand,
gives a dipole moment of 2.65 Debye\cite{burnham02a}, a significantly lower value.

Finally, the average potential energy of the 
liquid 
predicted by the SCME potential and classical trajectory calculations
is -10.8 kcal/mol per molecule. 
The best experimental estimate is
-9.86 kcal/mol per molecule\cite{dorsey40} but there quantum mechanical, zero point energy effects are included. 
In order to obtain a closer comparison with experiments, one should
employ a quantum mechanical simulation to properly take into account zero point energy effects
since the SCME is derived to reproduce the potential energy surface without any quantum corrections.
Model potentials parametrized to reproduce experimental
properties give results that are closer to experiment. For example, TIP4P predicts
an average energy of -9.83 kcal/mol per molecule\cite{burnham02a}, while the closely related
TIP4P-FQ potential gives a value of -9.92 kcal/mol per molecule\cite{chialvo00}.

\subsection{Ice}
\label{subsec:ice}

\begin{figure}
\includegraphics[scale=0.20]{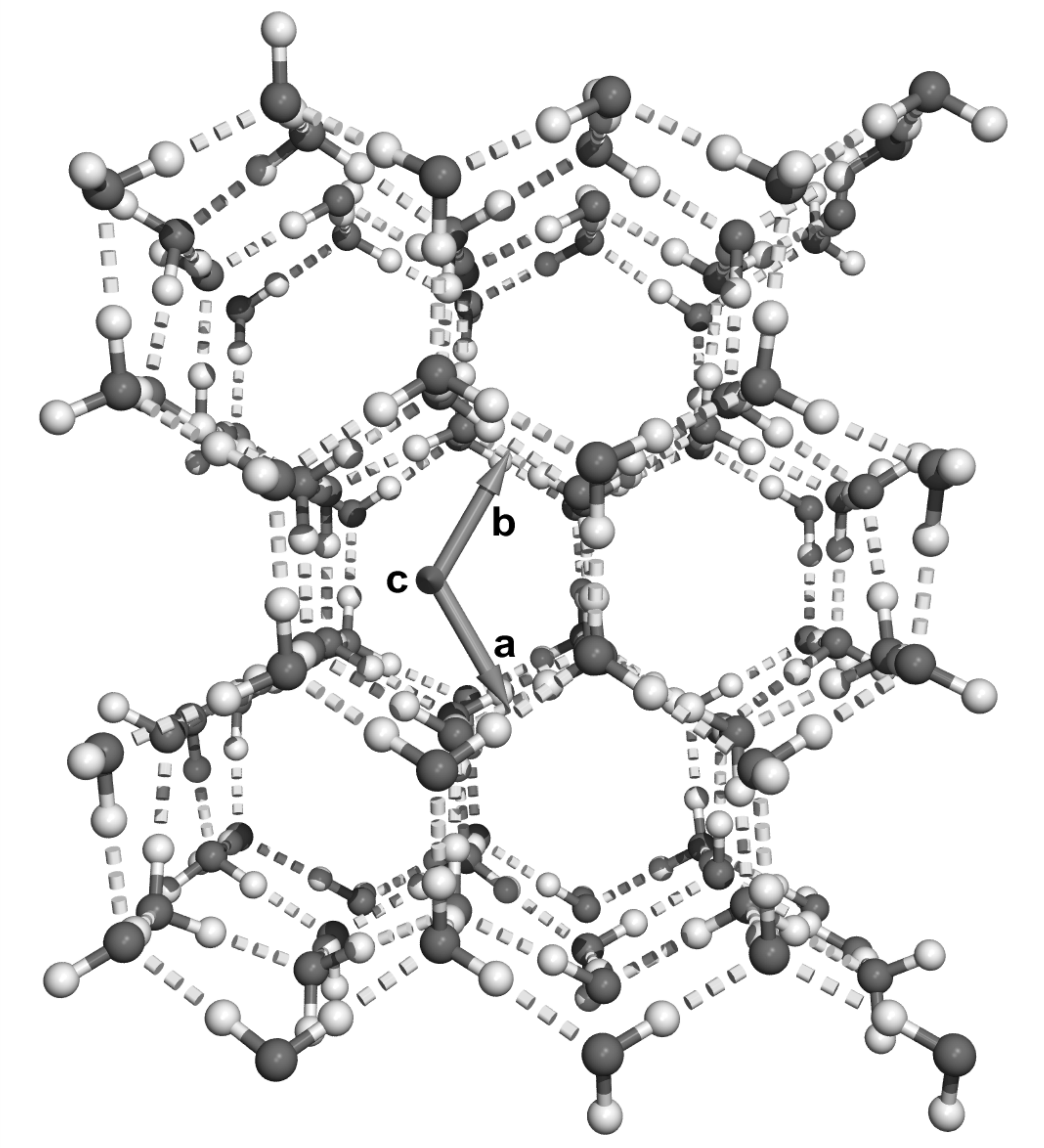}
\caption{\label{fig:ice_ih_cell}
         Typical orthorombic cell used for the simulation of
         (proton disordered) ice Ih.
        }
\end{figure}

One of our main interests in developing this
new potential function is to use it for simulations of ice growth.
The present study is limited
to the most common phase of crystalline ice, i.e. ice Ih. 
We simulated a
crystal sample containing 96 water molecules, built from $3\times2\times2$
repetitions of a generic 8-molecule orthogonal cell\cite{petrenko99b}.
Since ice Ih is proton-disordered, a Monte Carlo algorithm was used to generate ten
different cells that comply with the ice
rules and have null overall dipole moments. Figure \ref{fig:ice_ih_cell} shows
a typical example of the cells used in this work. As in the case of liquid water, the
properties discussed in this section were averaged over the different cells
used. Table \ref{tab:iceihgeom} presents the energy, conformational parameters
and electric properties of ice Ih. The values obtained with the SCME
potential are compared with experimental, density functional and model potential results
when available.

The many-body component of the repulsion energy 
in the SCME interaction potential function was adjusted to fit both the MP2 dimer potential
energy surface and the experimental cohesive energy and lattice parameters
of ice\cite{petrenko99c},
after the cohesive energy had been corrected to remove thermal and zero-point energy effects. 
As a result, the cohesive energy for ice is better reproduced by SCME than
for example the PW91 density functional\cite{hamann97} and several
pairwise additive and polarizable potentials. 
SCME also gives good agreement with the experimental
lattice parameters, the calculated values being only slightly
smaller (by $\sim$0.03 \mbox{\AA}). 
This small error, nevertheless, makes the density slightly too high.
The value predicted by SCME is, however, a significant
improvement over simple pair potentials such as TIP4P.

The average O-O distance and
the bulk modulus are useful measures of the quality of the potential since these
were not included in the fitting of the repulsive component.
The value predicted for the former is 2.742 \mbox{\AA}, only 0.01 \AA\ smaller than the
experimental value. This small overbinding is related to the underestimation
of the lattice parameters discussed above. In the case of the bulk modulus, the SCME
value is better than that obtained with DFT and significantly better
than those from other model potentials. It is, however, somewhat larger than the
experimentally determined value, making the potential slightly too stiff.

Also included in Table \ref{tab:iceihgeom} is the dipole moment of the monomer embedded
in the ice Ih lattice. As discussed in the previous section, the definition of the
dipole moment in ice is ambiguous and both the experimental\cite{whalley78} and
theoretical values present in the literature cover a rather wide range\cite{batista99a}.
The multipole expansion on which the SCME potential is based gives a value that is larger than many previous estimates, even by as much as
0.5 Debye\cite{note2}.


\section{Conclusions}
\label{sec:concl}

We have presented and tested a new model potential for the interaction between
water molecules based on a single-center multipole expansion (SCME) up to and including
hexadecapole and including both dipole and quadrupole polarizability.
Since point charges are not included, it is possible in some cases to simply truncate the potential at long range and thereby avoid the evaluation 
of Ewald sums. This reduces the computational effort significantly and while this potential function has many terms and a detailed description of
the electrostatics through a multipole expansion, it is still computationally efficient and applicable to large and complex systems.
The electrostatic, induction and dispersion components of the energy are obtained from
\textit{ab initio} and experimental molecular properties of the monomer, while the 
repulsive part of the potential was adjusted to reproduce \textit{ab initio} results 
for the dimer and the small ring-shaped clusters as well as the experimentally 
determined cohesive energy of ice Ih.
Since the electrostatics are evaluated including both dipole and quadrupole polarization through a self-consistency procedure, 
the potential should be transferable to a wide range of systems, well beyond the few that were used in the parametrization.

Our test results showed that, in general, the SCME potential is equally or even more
accurate than other sophisticated model potentials currently available.
The binding
energy and structure of small clusters are in quite good agreement
with the best available theoretical estimates. Some of the more subtle features of
the potential energy surface of the water dimer are well reproduced.
With the exception of the water trimer, the interaction energy for the ring clusters
are in excellent agreement with MP2/CBS results. For other clusters, such as the most
stable isomers of the water hexamer, the absolute values of the interaction energy is
less accurate, but the relative values for the different conformers are in good agreement with best estimates, such as CCSD(T) calculations.
In the case of the condensed phases, the energy and structural parameters are in
excellent agreement with experiment. SCME reproduces the radial distribution function
curves of the liquid and the lattice structure of ice Ih quite well.

The systematic deviations observed for the (H$_2$O)$_{n=2-6}$ clusters show
that there is still room for improvement. In particular, the structure obtained
for the water trimer could be improved. We believe these problems
originate mostly from the lack of flexibility of the functional form used for the 
repulsive
exchange interaction. Other sources of error can probably be found in the damping
function used for the electric fields and possibly also in the values used for the
multipole moments and polarizabilities.
However, the functional form used in SCME includes the essential physics of the problem and 
it should be possible to obtain a highly accurate parametrization of the water interaction with this form
using a more systematic parametrization from high level {\it ab initio} calculations.


\begin{thebibliography}{100}

\bibitem{ball2001}
Ball P, \emph{Life's matrix: a biography of water},  (University of California
  Press2001)

\bibitem{cheung2002}
Cheung M~S, Garc{\'\i}a A~E and Onuchic J~N, 2002 \emph{Proc. Natl. Acad. Sci.
  (USA)}, \textbf{99} 685

\bibitem{li2011}
Li I~T and Walker G~C, 2011 \emph{Pro. Natl. Acad. Sci. (USA)}, \textbf{108}
  16527--16532

\bibitem{snyder2011}
Snyder P~W, Mecinovi{\'c} J, Moustakas D~T, Thomas~III S~W, Harder M, Mack E~T,
  Lockett M~R, H{\'e}roux A, Sherman W and Whitesides G~M, 2011 \emph{Proc.
  Natl. Acad. Sci. (USA)}, \textbf{108} 17889--17894

\bibitem{grossman2011}
Grossman M, Born B, Heyden M, Tworowski D, Fields G~B, Sagi I and Havenith M,
  2011 \emph{Nature Struct. \& Mol. Bio.}, \textbf{18} 1102--1108

\bibitem{speedy1976}
Speedy R~J and Angell C~A, 1976 \emph{The Journal of Chemical Physics},
  \textbf{65(3)} 851--858

\bibitem{poole1992}
Poole P~H, Sciortino F, Essmann U and Stanley H~~E, 1992 \emph{Nature},
  \textbf{360(6402)} 324--328

\bibitem{debenedetti2003}
Debenedetti P~G, 2003 \emph{J. Phys: Cond. Matt.}, \textbf{15(45)} R1669

\bibitem{pierrehumbert99}
Pierrehumbert R~T.
\newblock In Clark P~U, Webb R~S and Keigwin L~D (eds.) \emph{Mechanisms of
  Global Change at Millennial Timescales}, volume 112 of \emph{Geophys. Monogr.
  Ser.},  (American Geophysical Union, Washington DC1999)

\bibitem{held00}
Held I~M and Soden B~J, 2000 \emph{Annu. Rev. Energ. Environ.}, \textbf{25} 441

\bibitem{petrenko99a}
Petrenko V~F and Withworth R~W, \emph{Physics of Ice},  (Oxford University
  Press1999)

\bibitem{smith02}
Smith J, Stone R and Fahrenkamp-Uppenbrink J, 2002 \emph{Science}, \textbf{297}
  1489

\bibitem{paterson94}
Paterson W~S~B, \emph{The Physics of Glaciers},  (Pergamon/Elsevier Science,
  Oxford1994)

\bibitem{pruppacher97}
Pruppacher H~R and Klett J~D, \emph{Microphysics of Clouds and Precipitation},
  (Kluwer Academic Publishers, Dordrecht1997)

\bibitem{ehrenfreund02}
Ehrenfreund P, Composition of comets and interstellar dust.
\newblock In Rickman H (ed.) \emph{Highlights of Astronomy}, volume~12 of
  \emph{IAU Symposia}, p. 229,  (Astronomical Soc. Pacific, San Fransisco2002)

\bibitem{ehrenfreund00}
Ehrenfreund P and Schutte W~A, Infrared observations of interstellar ices.
\newblock In Minh Y~C and van Dishoeck E~F (eds.) \emph{Astrochemistry: From
  Molecular Clouds to Planetary Systems}, volume 197 of \emph{IAU Symposia}, p.
  135,  (Astronomical Soc. Pacific, San Fransisco2000)

\bibitem{caro02}
Caro G~M~M, Meierhenrich U~J, Schutte W~A, Barbier B, Segovia A~A, Rosenbauer
  H, Thiemann W~H~P, Brack A and Greenberg J~M, 2002 \emph{Nature},
  \textbf{416} 403

\bibitem{burton35}
Burton E~F and Oliver W~F, 1935 \emph{Proc. R. Soc. Lon. A}, \textbf{153} 166

\bibitem{mayer82}
Mayer E and Br{\"{u}}ggeller P, 1982 \emph{Nature}, \textbf{298} 715

\bibitem{mccammon1985}
McCammon D, Moseley S, Mather J, Musholzky R, Fiorini E and Niinikoski T, 1985
  \emph{Nature}, \textbf{314} 7

\bibitem{jenniskens94}
Jenniskens P and Blake D~F, 1994 \emph{Science}, \textbf{265} 753

\bibitem{hallbrucker89}
Hallbrucker A, Mayer E and Johari G~P, 1989 \emph{J. Phys. Chem.}, \textbf{93}
  4986

\bibitem{dash95}
Dash J~G, Fu H and Wettlaufer J~S, 1995 \emph{Reports on Progress in Physics},
  \textbf{58} 115

\bibitem{furukawa97}
Furukawa Y and Nada H, 1997 \emph{J. Phys. Chem. B}, \textbf{101} 6167

\bibitem{kroes92}
Kroes G~J, 1992 \emph{Surf. Sci.}, \textbf{275} 365

\bibitem{morgenstern97}
Morgenstern M, M{\"{u}}ller J, Michely T and Comsa G, 1997 \emph{Z. Phys.
  Chem.}, \textbf{198} 43

\bibitem{materer97}
Materer N, Starke U, Barbieri A, Hove M~A~V, Somorjai G~A, Kroes G~J and Minot
  C, 1997 \emph{Surf. Sci.}, \textbf{381} 190

\bibitem{braun98}
Braun J, Glebov A, Graham A~P, Menzel A and Toennies J~P, 1998 \emph{Phys. Rev.
  Lett.}, \textbf{80} 2638

\bibitem{kryachko90}
Kryachko E~S and Lude{\~{n}}a E~V, \emph{Energy Density Functional Theory of
  Many-Electron Systems},  (Kluwer Academic Publishers, Dordrecht1990)

\bibitem{march96}
March N~H, \emph{Electron Correlation in Molecules and Condensed Phases}.
\newblock Physics of Solids and Liquids,  (Plenum Press, New York1996)

\bibitem{frenkel02}
Frenkel D and Smit B, \emph{Understanding Molecular Simulation},  (Academic
  Press, San Diego2002)

\bibitem{stone96a}
Stone A~J, \emph{The Theory of Intermolecular Forces}, pp. 74--75,  (Clarendon
  Press, Oxford1996)

\bibitem{xantheas95}
Xantheas S~S, 1995 \emph{J. Chem. Phys.}, \textbf{102} 4505

\bibitem{delbene95}
Bene J~E~D, Person W~B and Szczepaniak K, 1995 \emph{J. Phys. Chem.},
  \textbf{99} 10705

\bibitem{sprik96}
Sprik M, Hutter J and Parrinello M, 1996 \emph{J. Chem. Phys.}, \textbf{105}
  1142

\bibitem{grossman04}
Grossman J~C, Schwegler E, Draeger E~W, Gygi F and Galli G, 2004 \emph{J. Chem.
  Phys.}, \textbf{120} 300

\bibitem{ireta2004}
Ireta J, Neugebauer J and Scheffler M, 2004 \emph{J. Phys. Chem. A},
  \textbf{108(26)} 5692--5698

\bibitem{santra2007}
Santra B, Michaelides A and Scheffler M, 2007 \emph{J. Chem. Phys.},
  \textbf{127} 184104

\bibitem{perez95}
P{\'{e}}rez-Jord{\'{a}} J~M and Becke A~D, 1995 \emph{Chem. Phys. Lett.},
  \textbf{233} 134

\bibitem{vanmourik02}
van Mourik T and Gdanitz R~J, 2002 \emph{J. Chem. Phys.}, \textbf{116} 9620

\bibitem{dion04}
Dion M, Rydberg H, Schr{\"o}der E, Langreth D~C and Lundqvist B~I, 2004
  \emph{Phys. Rev. Lett.}, \textbf{92} 246401

\bibitem{vydrov09}
Vydrov O~A and Voorhis T~V, 2009 \emph{Phys. Rev. Lett.}, \textbf{103} 63004

\bibitem{lee10}
Lee K, Murray E~D, Kong L, Lundqvist B~I and Langreth D~C, 2010 \emph{Phys.
  Rev. B}, \textbf{82} 081101

\bibitem{klimes10}
Klime{\v{s}} J, Bowler D~R and Michaelides A, 2010 \emph{J. Phys.: Cond.
  Matter}, \textbf{22} 022201

\bibitem{wallqvist99}
Wallqvist A and Mountain R~D, Molecular models of water: Derivation and
  description.
\newblock volume~13 of \emph{Reviews in Computational Chemistry}, p. 183,
  (Wiley-VCH, New York1999)

\bibitem{vega11}
Vega C and Abascal J~L~F, 2011 \emph{Phys. Chem. Chem. Phys.}, \textbf{13}
  19663--19688

\bibitem{berendsen87}
Berendsen H~J~C, Grigera J~R and Straatsma T, 1987 \emph{J. Phys. Chem.},
  \textbf{87} 6269

\bibitem{jorgensen83}
Jorgensen W~L, Chandrasekhar J, Madura J~D, Impey R~W and Klein M~L, 1983
  \emph{J. Chem. Phys.}, \textbf{79} 926

\bibitem{horn04}
Horn H~W, Swope W~C, Pitera J~W, Madura J~D, Dick T~J, Hura G~L and Head-Gordon
  T, 2004 \emph{J. Chem. Phys.}, \textbf{120} 9665

\bibitem{abascal05}
Abascal J~L~F and Vega C, 2005 \emph{J. Chem. Phys.}, \textbf{123} 234505

\bibitem{pedulla98}
Pedulla J~M and Jordan K~D, 1998 \emph{Chem. Phys.}, \textbf{239} 593

\bibitem{mhin93}
Mhin B~J, Kim J~S, Lee S and Kim K~S, 1993 \emph{J. Chem. Phys.}, \textbf{100}
  4484

\bibitem{wales98}
Wales D~J and Hodges M~P, 1998 \emph{Chem. Phys. Lett.}, \textbf{286} 65

\bibitem{dang97}
Dang L~X and Chang T~M, 1997 \emph{J. Chem. Phys.}, \textbf{106} 8149

\bibitem{niesar90}
Niesar U, Corongiu G, Clementi E, Kneller G~R and Bhattacharya D~K, 1990
  \emph{J. Phys. Chem.}, \textbf{94} 7949

\bibitem{dong01}
Dong S, Wang Y and Li J, 2001 \emph{Chem. Phys.}, \textbf{270} 309

\bibitem{millot92}
Millot C and Stone A~J, 1992 \emph{Mol. Phys.}, \textbf{77} 439

\bibitem{millot98}
Millot C, Soetens J~C, Costa M~T~C~M, Hodges M~P and Stone A~J, 1998 \emph{J.
  Phys. Chem. A}, \textbf{102} 754

\bibitem{burnham02a}
Burnham C~J and Xantheas S~S, 2002 \emph{J. Chem. Phys.}, \textbf{116} 1500

\bibitem{burnham02b}
Burnham C~J and Xantheas S~S, 2002 \emph{J. Chem. Phys.}, \textbf{116} 5115

\bibitem{fanourgakis08}
Fanourgakis G~S and Xantheas S~S, 2008 \emph{J. Chem. Phys.}, \textbf{128}
  074506

\bibitem{frenkel02a}
Frenkel D and Smit B, \emph{Understanding Molecular Simulation}, pp. 292--306,
  (Academic Press, San Diego2002)

\bibitem{ewald21}
Ewald P, 1921 \emph{Ann. Phys.}, \textbf{64} 253

\bibitem{allen87a}
Allen M~P and Tildesley D~J, \emph{Computer Simulations of Liquids}, pp.
  155--162,  (Clarendon Press, Oxford1987)

\bibitem{ichiye96}
Liu Y and Ichiye T, 1996 \emph{J. Phys. Chem.}, \textbf{100} 2723

\bibitem{barnes79}
Barnes P, Finney J~L, Nicholas J~D and Quinn J~E, 1979 \emph{Nature},
  \textbf{282} 459

\bibitem{batista98}
Batista E~R, Xantheas S~S and J{\'{o}}nsson H, 1998 \emph{J. Chem. Phys.},
  \textbf{109} 4546

\bibitem{batista00}
Batista E~R, Xantheas S~S and J{\'{o}}nsson H, 2000 \emph{J. Chem. Phys.},
  \textbf{112} 3285

\bibitem{stone96b}
Stone A~J, \emph{The Theory of Intermolecular Forces}, pp. 50--63,79--104,
  (Clarendon Press, Oxford1996)

\bibitem{ahlrichs77}
Ahlrichs R, Penco P and Scoles G, 1977 \emph{Chem. Phys.}, \textbf{19} 119

\bibitem{stone96c}
Stone A~J, \emph{The Theory of Intermolecular Forces}, p. 118,  (Clarendon
  Press, Oxford1996)

\bibitem{stone96}
Stone A~J, \emph{The Theory of Intermolecular Forces},  (Clarendon Press,
  Oxford1996)

\bibitem{dyke73}
Dyke T and Muenter J, 1973 \emph{J. Chem. Phys.}, \textbf{59} 3125

\bibitem{verhoeven70}
Verhoeven J and Dymanus A, 1970 \emph{J. Chem. Phys.}, \textbf{52} 3222

\bibitem{gray84}
Gray C~G and Gubbins K~E, \emph{Theory of Molecular Fluids},  (Clarendon Press,
  Oxford1984)

\bibitem{stone96d}
Stone A~J, \emph{The Theory of Intermolecular Forces}, pp. 94--96,  (Clarendon
  Press, Oxford1996)

\bibitem{tang84}
Tang K~T and Toennies J~P, 1984 \emph{J. Chem. Phys.}, \textbf{80} 3726

\bibitem{wormer1992}
Wormer P~E~S and Hettema H, 1992 \emph{J. Chem. Phys.}, \textbf{97} 5592

\bibitem{COonPt1}
J{\'{o}}nsson H, Levi A~C and Weare J~H, 1984 \emph{Phys. Rev. B}, \textbf{30}
  2241

\bibitem{COonPt2}
J{\'{o}}nsson H, Levi A~C and Weare J~H, 1984 \emph{Surf. Sci.}, \textbf{148}
  126

\bibitem{stone97}
Stone A~J, Dullweber A, Hodges M~P, Popelier P and Wales D~J, 1997.
\newblock Orient 3.2j

\bibitem{frisch98}
Frisch M~J \emph{et~al.}, 1998.
\newblock Gaussian 98, revision a.7

\bibitem{burnham02c}
Burnham C~J, Xantheas S~S and Harrison R~J, 2002 \emph{J. Chem. Phys.},
  \textbf{116} 1493

\bibitem{Andersen80}
Andersen H, 1980 \emph{J. Chem. Phys.}, \textbf{72} 2384

\bibitem{skinner13}
{Skinner, L B and Huang, C and Schlesinger, D and Pettersson, L G M and
  Nilsson, A and Benmore, C J}, 2013 \emph{J. Chem. Phys.}, \textbf{138} 074506

\bibitem{soper07}
Soper A~K, 2007 \emph{J. Phys.: Cond. Matt.}, \textbf{19} 335206

\bibitem{wikfeldt09}
Wikfeldt K~T, Leetmaa M, Ljungberg M~P, Nilsson A and Pettersson L~G~M, 2009
  \emph{J. Phys. Chem. B}, \textbf{113} 6246--6255

\bibitem{kuharski84}
Kuharski R~A and Rossky P~J, 1984 \emph{Chem. Phys. Lett.}, \textbf{103} 357

\bibitem{lobaugh97}
Lobaugh J and Voth G, 1997 \emph{J. Chem. Phys.}, \textbf{106} 2400

\bibitem{morrone08}
Morrone J and Car R, 2008 \emph{Phys. Rev. Lett.}, \textbf{101} 17801

\bibitem{batista99a}
Batista E~R, Xantheas S~S and J{\'{o}}nsson H, 1999 \emph{J. Chem. Phys.},
  \textbf{111} 6011

\bibitem{silvestrelli99}
Silvestrelli P~L and Parrinello M, 1999 \emph{J. Chem. Phys.}, \textbf{111}
  3572

\bibitem{dorsey40}
Dorsey N~E, \emph{Properties of Ordinary Water-Substance in All its Phases:
  water-vapor, water and all the ices}, volume~81 of \emph{ACS Monograph
  Series},  (Reinhold Publishing Corporation, New York1940)

\bibitem{chialvo00}
Chialvo A~A, Yezdimer E, Driesner T, Cummings P~T and Simonson J~M, 2000
  \emph{Chem. Phys.}, \textbf{258} 109

\bibitem{petrenko99b}
Petrenko V~F and Withworth R~W, \emph{Physics of Ice}, p.~21,  (Oxford
  University Press1999)

\bibitem{petrenko99c}
Petrenko V~F and Withworth R~W, \emph{Physics of Ice}, pp. 29--30,  (Oxford
  University Press1999)

\bibitem{hamann97}
Hamann D~R, 1997 \emph{Phys. Rev. B}, \textbf{55} R10157

\bibitem{whalley78}
Whalley E, 1978 \emph{J. Glaciology}, \textbf{21} 13

\bibitem{note2}
It should be noticed that the model presented in Ref.\cite{batista98} used
  values for the polarization tensors calculated with respect to the Oxygen
  atom without transforming them to a center of mass origin. As a result, the
  average molecular dipole moment for ice Ih presented in this work is larger
  than the estimate reported in Ref. \cite{batista98}.

\bibitem{hobbs74}
Hobbs P~V, \emph{Ice Physics},  (Clarendon Press, Oxford1974)

\bibitem{jenkins01}
Jenkins S and Morrison I, 2001 \emph{J. Phys.: Condens. Matter}, \textbf{13}
  9207

\bibitem{batista99b}
Batista E~R, 1999 \emph{Development of a New Water-Water Interaction Potential
  and Application to Molecular Processes in Ice}.
\newblock Ph.D. thesis, University of Washington

\bibitem{reimers82}
Reimers J~R, Watts R~O and Klein M~L, 1982 \emph{Chem. Phys.}, \textbf{64} 95

\end{thebibliography}

\pagebreak


\begin{table*}
\caption{ \label{tab:multipoles}
     Multipole moments of the water molecule used in the calculation of the
     electrostatic and induction components of the interaction energy.
     All values in atomic units. All moments defined with respect to the center of mass.
     Molecular orientation as shown in Figure \ref{fig:w1axes}.
        }
\begin{indented}
\item[]
\begin{tabular}{@{}lll} 

\br
\thl{Multipole Moment} &
\thl{Component}   &          \\
\mr
Dipole$^a$       & $\mu_{z}$         &  -0.72981\\
Quadrupole$^b$   & $\Theta_{xx}$     &   1.95532\\
                             & $\Theta_{yy}$     &  -1.85867\\
                             & $\Theta_{zz}$     &  -0.09665\\
Octupole$^c$     & $\Omega_{xxz}$    &  -3.27190\\
                             & $\Omega_{yyz}$    &   1.36606\\
                             & $\Omega_{zzz}$    &   1.90585\\
Hexadecapole$^c$ & $\Phi_{xxxx}$     &  -0.94903\\
                             & $\Phi_{xxyy}$     &  -3.38490\\
                             & $\Phi_{xxzz}$     &   4.33393\\
                             & $\Phi_{yyyy}$     &   4.09835\\
                             & $\Phi_{yyzz}$     &  -0.71345\\
                             & $\Phi_{zzzz}$     &  -3.62048\\
\br
\end{tabular}\\
{$^a$ \footnotesize{From Ref. \cite{dyke73}}} \\
{$^b$ \footnotesize From Ref. \cite{verhoeven70}}\\
{$^c$ \footnotesize From Ref. \cite{batista98}}                
\end{indented}
\end{table*}

\begin{table*}
\caption{\label{tab:polarizabilities} 
  Polarizabilities used in the calculation of the induction component
  of the interaction energy. All values in atomic units. All moments defined with respect to 
  the center of mass. Molecular orientation as shown in Figure \ref{fig:w1axes}.}
\begin{indented}
\item[]
\begin{tabular}{@{}lll} 
\br
\thl{Polarizability} &
\thl{Component}   &          \\
\mr
Dipole-Dipole$^a$         & $\alpha_{xx}$    &  10.31146\\
                                      & $\alpha_{yy}$    &   9.54890\\
                                      & $\alpha_{zz}$    &   9.90656\\
Dipole-Quadrupole$^a$     & $A_{x,xz}$       &  -8.42037\\
                                      & $A_{y,yz}$       &  -1.33400\\
                                      & $A_{z,xx}$       &  -2.91254\\
                                      & $A_{z,yy}$       &   4.72407\\
                                      & $A_{z,zz}$       &  -1.81153\\
Quadrupole-Quadrupole$^a$ & $C_{xx,xx}$      &  12.11907\\
                                      & $C_{xx,yy}$      &  -6.95326\\
                                      & $C_{xx,zz}$      &  -5.16582\\
                                      & $C_{xy,xy}$      &   7.86225\\
                                      & $C_{xz,xz}$      &  11.98862\\
                                      & $C_{yy,yy}$      &  11.24741\\
                                      & $C_{yy,zz}$      &  -4.29415\\
                                      & $C_{yz,yz}$      &   6.77226\\
                                      & $C_{zz,zz}$      &   9.45997\\
\br
\end{tabular}\\
$^a$ {\footnotesize These values correspond to a translational transformation of those reported in Ref. \cite{millot92}.}
\end{indented}
\end{table*}

\begin{table*}
\caption{ \label{tab:othpar} 
  Other parameters used in the calculation of the interaction energy. All values in atomic units.}
\begin{indented}
\item[]
\begin{tabular}{@{}lll} 
\br
\thl{Component}         & \thl{Parameter}  &       \\
\mr
Damping                 & $\tau_d $   &     2.32837906               \\
Electrostatic+Induction & $r_{h_1}$ &     9.44863332               \\
                        & $r_{l_2}$ &    17.00753997               \\
                        & $r_{h_2}$ &    20.78699330               \\
Dispersion $^a$              & $C_6$      &    46.44309964               \\
                        & $C_8$      &  1141.70326668               \\
                        & $C_{10}$  & 33441.11892923               \\
Repulsion               & $A$       &  1857.45898793               \\
                        & $C$       &     1.68708507 \mbox{$\times 10^{  6}$} \\
                        & $b$       &     1.44350000               \\
                        & $c$       &     1.83402715               \\
                        & $d$       &     0.35278471               \\
                        & $a_0$     &     1.02508535 \mbox{$\times 10^{ -1}$} \\
                        & $a_1$     &    -1.72461186 \mbox{$\times 10^{ -4}$} \\
                        & $a_2$     &     1.02195556 \mbox{$\times 10^{ -7}$} \\
                        & $a_3$     &    -2.60877107 \mbox{$\times 10^{-11}$} \\
                        & $a_4$     &     3.06054306 \mbox{$\times 10^{-15}$} \\
                        & $a_5$     &    -1.32901339 \mbox{$\times 10^{-19}$} \\
\br
\end{tabular}\\
$^a${\footnotesize From Ref. \cite{wormer1992}}
\end{indented}
\end{table*}

\begin{table*}
\caption{ \label{tab:w2geom}
         Comparison of the optimal structure of the water dimer obtained
         with different methods. See Figure \ref{fig:w2geom} for a definition
         of each structure coordinate.
}

\begin{indented}
\item[]
\begin{tabular}{@{}llllll} 
\br
\thl{Coordinate}         & \thl{Atoms} & \thc{MP2} & \thc{SCME} & \thc{NCC} & \thc{ASP-W4}\\
\mr
Distance [\AA] &  (1,2)      &   2.907   &   2.906   &   2.965   &   2.974\\
Angle [deg]    &  (1,a,2)    & 171.57    & 175.42    & 179.49    &-176.95\\
                         &  (1,2,X)    & 123.09    & 113.99    & 152.77    & 123.03\\
Dihedral [deg] &  (A,1,2,B)  & 122.96    & 125.27    & 109.50    & 122.98\\
\br
\end{tabular}
\end{indented}
\end{table*}

\begin{table*}
\caption{ \label{tab:w3geom}
         Comparison of the optimal structure of the water trimer obtained
         with different methods. See Figure \ref{fig:w3-6geom} for a definition
         of each structure coordinate.
}

\begin{indented}
\item[]
\begin{tabular}{@{}llllll} 
\br
\thl{Coordinate} & \thl{Atoms} & \thc{MP2} & \thc{SCME} & \thc{NCC} & \thc{ASP-W4}\\
\mr
Distance [\AA]       & (1,2)       &   2.799   &   2.840   &   2.865   &  2.868\\
            & (2,3)       &   2.798   &   2.843   &   2.868   &  2.865\\
            & (3,1)       &   2.800   &   2.858   &   2.871   &  2.884\\
Angle [deg]       & (1,a,2)     &  151.26   &  158.73   &  149.28   & 148.47\\
            & (2,b,3)     &  151.11   &  158.65   &  149.13   & 148.24\\
            & (3,c,1)     &  148.39   &  157.22   &  147.99   & 145.70\\
Dihedral [deg]    & (A,1,2,3)   & -129.13   & -114.08   & -145.80   &-120.67\\
            & (B,2,3,1)   &  118.38   &  110.63   &  128.46   & 119.20\\
            & (C,3,1,2)   & -122.71   & -111.14   & -140.01   &-121.92\\
            & (A,1,2,B)   &  129.49   &  148.63   &  114.99   & 144.04\\
            & (B,2,3,C)   & -133.86   & -151.90   & -130.35   &-135.34\\
            & (C,3,1,A)   &  -21.87   &  -14.78   &  -36.55   & -27.72\\
\br
\end{tabular}
\end{indented}
\end{table*}

\begin{table*}
\caption{ \label{tab:w4geom}
         Comparison of the optimal structure of the water tetramer obtained
         with different methods. See Figure \ref{fig:w3-6geom} for a definition
         of each structure coordinate.
}

\begin{indented}
\item[]
\begin{tabular}{@{}llllll} 
\br
\thl{Coordinate} & \thl{Atoms} & \thc{MP2} & \thc{SCME} & \thc{NCC} & \thc{ASP-W4}\\
\mr
Distance [\AA]       & (1,2)       &   2.743   &   2.737   &   2.822   &   2.844\\
Angle [deg]       & (1,a,2)     &  167.64   &  173.27   &  165.69   &  163.69\\
Dihedral [deg]    & (4,1,2,3)   &   -0.48   &    1.35   &   -9.90   &    1.75\\
            & (A,1,2,3)   &  123.45   &  118.24   &  128.17   &  123.71\\
            & (A,1,2,B)   & -123.69   & -134.93   & -113.86   & -132.36\\
\br
\end{tabular}
\end{indented}
\end{table*}

\begin{table*}
\caption{ \label{tab:w5geom}
         Comparison of the optimal structure of the water pentamer obtained
         with different methods. See Figure \ref{fig:w3-6geom} for a definition
         of each structure coordinate.
}
\begin{indented}
\item[]
\begin{tabular}{@{}llllll} 
\br
\thl{Coordinate} & \thl{Atoms} & \thc{MP2} & \thc{SCME} & \thc{NCC} & \thc{ASP-W4}\\
\mr
Distance [\AA]       & (1,2)       &   2.722   &   2.717   &   2.806   &   2.839\\
            & (2,3)       &   2.725   &   2.719   &   2.809   &   2.843\\
            & (3,4)       &   2.734   &   2.729   &   2.815   &   2.869\\
            & (4,5)       &   2.726   &   2.716   &   2.810   &   2.840\\
            & (5,1)       &   2.723   &   2.717   &   2.807   &   2.838\\
Angle [deg]       & (1,a,2)     &  175.91   &  177.41   &  173.57   &  168.74\\
            & (2,b,3)     &  176.77   &  178.11   &  174.14   &  169.26\\
            & (3,c,4)     &  173.01   &  176.99   &  172.94   &  174.42\\
            & (4,d,5)     &  176.65   &  178.16   &  175.72   &  169.12\\
            & (5,e,1)     &  175.72   &  177.07   &  173.08   &  168.69\\
Dihedral [deg]    & (1,2,3,4)   &   15.23   &   10.69   &   19.02   &    2.84\\
            & (2,3,4,5)   &   -9.19   &  -11.36   &   -4.26   &   -6.29\\
            & (3,4,5,1)   &   -0.28   &    7.70   &  -12.10   &    7.43\\
            & (4,5,1,2)   &    9.66   &   -1.07   &   23.79   &   -5.66\\
            & (5,1,2,3)   &  -15.46   &   -5.99   &  -26.54   &    1.70\\
            & (A,1,2,3)   &  114.41   &  115.80   &  116.96   &  124.04\\
            & (B,2,3,4)   & -113.02   & -111.00   & -123.06   & -119.48\\
            & (C,3,4,5)   &  117.47   &  107.62   &  139.54   &  117.55\\
            & (D,4,5,1)   &  136.05   &  134.38   &  159.73   &  127.55\\
            & (E,5,1,2)   & -115.38   & -119.66   & -118.00   & -126.55\\
            & (A,1,2,B)   & -124.93   & -129.72   & -114.85   & -126.16\\
            & (B,2,3,C)   &  124.95   &  136.35   &  106.87   &  125.84\\
            & (C,3,4,D)   &   -8.70   &   -9.41   &  -27.07   &    9.43\\
            & (D,4,5,E)   & -106.47   & -113.30   &  -72.81   & -123.28\\
            & (E,5,1,A)   &  123.43   &  126.14   &  112.86   &  124.16\\
\br
\end{tabular}
\end{indented}
\end{table*}

\begin{table*}
\caption{ \label{tab:w6geom}
         Comparison of the optimal structure of the water hexamer obtained
         with different methods. See Figure \ref{fig:w3-6geom} for a definition
         of each structure coordinate.
}
\begin{indented}
\item[]
\begin{tabular}{@{}llllll} 
\br
\thl{Coordinate} & \thl{Atoms} & \thc{MP2} & \thc{SCME} & \thc{NCC} & \thc{ASP-W4}\\
\mr
Distance [\AA]       & (1,2)       &   2.716   &   2.728   &   2.804   &   2.837\\
Angle [deg]       & (1,a,2)     &  178.73   &  174.80   &  176.07   &  167.16\\
Dihedral [deg]    & (1,2,3,4)   &   20.63   &   12.90   &   35.16   &   -4.90\\
            & (A,1,2,3)   &  112.60   &  113.61   &  114.05   &  126.92\\
            & (A,1,2,B)   & -120.40   & -125.92   & -106.97   & -120.30\\
\br
\end{tabular}
\end{indented}
\end{table*}

\begin{table*}
\caption{ \label{tab:wnenergies}
         Comparison of the interaction energy of the ring shaped (H$_2$O)$_n$ clusters
         obtained with different methods. All values in kcal/mol.
}
\begin{indented}
\item[]
\begin{tabular}{@{}cclllll} 
\br
\thl{n} &\thl{MP2/CBS$^a$} & \thl{SCME} & \thl{NCC} & \thl{ASP-W4} & \thl{TTM2-R$^b$} \\
\mr
 2      &       -4.98  &  -5.11     &   -5.09   &   -4.97      &   -4.98      \\
 3      &      -15.8   & -16.22     &  -14.88   &  -15.28      &  -15.59      \\
 4      &      -27.6   & -28.95     &  -25.51   &  -26.19      &  -27.03      \\
 5      &      -36.3   & -37.88     &  -33.92   &  -33.82      &  -36.05      \\
 6      &      -44.8   & -45.86     &  -41.80   &  -41.68      &  -44.28      \\
\br
\end{tabular}\\
$^a$ {\footnotesize From Ref. \cite{burnham02c}}\\
$^b$ {\footnotesize From Ref. \cite{burnham02a}}
\end{indented}
\end{table*}

\begin{table*}
\caption{ \label{tab:w6othenerg}
         Comparison of the interaction energy of the cage, prism, book
         and ring isomers of (H$_2$O)$_6$ obtained with different methods.
         All values in kcal/mol.
}
\begin{indented}
\item[]
\begin{tabular}{@{}cclllll} 
\br
\thl{Conformation}& \thl{$\Delta$CCSD(T)/CBS$^a$} & \thl{SCME} & \thl{NCC} & \thl{ASP-W4} & \thl{TTM2-R$^b$} \\
\mr
Prism  &   -46.2       &     -47.56 &    -44.78 &    -45.87    &       -45.11 \\
Cage   &   -45.9       &     -47.64 &    -44.41 &    -44.74    &       -45.67 \\
Book   &   -45.5       &     -47.77 &    -43.11 &    -43.61    &       -45.14 \\
Ring   &   -44.5       &     -45.86 &    -41.80 &    -41.68    &       -44.28 \\
\br
\end{tabular}\\
$^a$ {\footnotesize From Ref. \cite{klimes10}}\\
$^b$ {\footnotesize From Ref. \cite{burnham02a}}
\end{indented}
\end{table*}

\begin{table}
\caption{ \label{tab:iceihgeom}
         Comparison between some experimental properties of ice Ih at 0K
         and those obtained with \textit{ab initio} methods, the SCME
         model potential and other potentials commonly used:
         $\Delta E_{Lattice}$ (lattice energy, in eV/molec),
         $\langle r_{OO} \rangle$ (average O-O distance, in \AA),
         $a$, $b$, $c$ (lattice parameters for the eight-molecule orthorhombic cell, in \AA),
         $\rho$ (density, in g/cm$^3$),
         $V_{molec}$ (molecular volume, in \AA$^3$/molec),
         $K$ (bulk modulus, in MPa),
         $\mu_{molec}$ (molecular dipole moment, in Debye).
}
\begin{tabular}{@{}lllllllll}
\br
\thl{Property}           &\thc{Exp.$^a$}&\thc{PW91$^b$}&\thc{SCME}          &\thc{TIP4P$^c$}&\thc{RWK2$^d$}&\thc{DC$^e$}&\thc{TTM2-R$^f$}\\
\mr
$\Delta E_{Lattice}$     &  -0.6110 &  -0.55$^g$  &-0.6109\mbox{$\pm$0.0049}             &  -0.634   &  -0.555  & -0.550 &  -.6370    \\
$\langle r_{OO} \rangle$ &   2.751  &   2.70  & 2.742\mbox{$\pm$0.004}              &   2.683   &          &  2.738 &            \\
$a$                      &  4.4969  &   4.41  & 4.470\mbox{$\pm$0.025}             &           &          &        &  4.478     \\
$b$                      &  7.7889  &   7.63  & 7.747\mbox{$\pm$0.052}             &           &          &        &  7.756     \\
$c$                      &  7.3211  &   7.20  & 7.287\mbox{$\pm$0.029}             &           &          &        &  7.314     \\
$\rho$                   &  0.933   &   0.989, 0.954$^g$ &0.948\mbox{$\pm$0.004}               &   1.009   &   0.942  &  0.960 &  0.942     \\
$V_{molec}$              & 32.05    &  30.3,  31.35$^g$ &31.55\mbox{$\pm$0.15}               &  29.62    &  31.73   & 31.14  & 31.75      \\
$K$                      & 10.9     &  13.5   &11.4\mbox{$\pm$0.3}                &  16.6     &  18.0    &        &            \\
$\mu_{molec}$            & 2.90     &   2.8   &3.50\mbox{$\pm$0.07}&   2.18    &          &  3.02  &  2.86$^h$      \\
\br
\end{tabular}\\

$^a$ {\footnotesize All values from Ref. \cite{petrenko99a} with the exception of the
                 bulk modulus, taken from Ref. \cite{hobbs74}.}\\
$^b$ {\footnotesize All values from Ref. \cite{jenkins01} unless indicated.}\\
$^c$ {\footnotesize All values from Ref. \cite{dong01} with the exception of the
                 bulk modulus, taken from Ref. \cite{batista99b}.}\\
$^d$ {\footnotesize From Ref. \cite{reimers82}.}\\
$^e$ {\footnotesize From Ref. \cite{dong01}.}\\
$^f$ {\footnotesize From Ref. \cite{burnham02a}.}\\
$^g$ {\footnotesize From Ref. \cite{hamann97}.}\\
$^h$ {\footnotesize Calculated at 100K.}

\end{table}


\clearpage

\end{document}